\documentclass[aps,prl,amssymb,groupedaddress]{revtex4}
\usepackage{epsfig}
\usepackage[english]{babel}

\def\a{\alpha}
\def\b{\beta}
\def\D{\Delta}
\def\d{\delta}
\def\e{\epsilon}
\def\G{\Gamma}
\def\g{\gamma}

\def\l{\lambda}
\def\n{\eta}

\def\u{\mu}

\def\Y{\Psi}

\def\hf{\frac{1}{2}}

\def\px{\approx}

\def\={&=&}

\def\({\left(}
\def\){\right)}
\def\[{\left[}
\def\]{\right]}
\def\<{\left\langle}
\def\>{\right\rangle}

\def\uk{{\bf \hat{k}}}
\def\un{{\bf \hat{n}}}

\def\ux{{\bf \hat{x}}}
\def\bk{{\bf k}}
\def\bn{{\bf n}}

\def\eq{\begin{eqnarray}}
\def\qe{\end{eqnarray}}
\def\bib{\bibitem}
\def\curl{\mathcal}
\def\it{\textit}
\def\tbf{\textbf}
\def\and{\quad \mbox{and} \quad}
\def\imp{\quad \Rightarrow \quad}
\newcommand{\rf}[1]{eq.(\ref{#1})}
\newcommand{\rfig}[1]{(see figure \ref{#1})}

\begin{document}

\title{Primordial non-Gaussianity and the CMB bispectrum}

\author{J.R.~Fergusson}

\author{E.P.S.~Shellard}

\affiliation{Department of Applied Mathematics and Theoretical Physics,
Centre for Mathematical Sciences,\\
University of Cambridge,
Wilberforce Road, Cambridge CB3 0WA, United Kingdom}

\date{\today}

\begin{abstract}

\noindent We present a new formalism, together with efficient numerical methods, to directly calculate the CMB bispectrum today from a given primordial bispectrum using the full linear radiation transfer functions.   Unlike previous analyses which have assumed simple separable ans\"atze for the bispectrum, this work applies to a primordial bispectrum of almost arbitrary functional form, for which there may have been both horizon-crossing and superhorizon contributions. We employ adaptive methods on a hierarchical triangular grid and we establish their accuracy by direct comparison with an exact analytic solution, valid on large angular scales. We demonstrate that we can calculate the full CMB bispectrum to greater than 1\% precision out to multipoles $l<1800$ on reasonable computational timescales. We plot the bispectrum for both the superhorizon (`local') and horizon-crossing (`equilateral') asymptotic limits, illustrating its oscillatory nature which is analogous to the CMB power spectrum. 

\end{abstract}

\pacs{1}

\maketitle

%%%%%%%%%%%%%%%%%%%%%%%%%%%%%%%%% INTRODUCTION %%%%%%%%%%%%%%%%%%%%%%%%%%%%%%%%

\section{Introduction}

Measurements of the cosmic microwave background (CMB) radiation are now strongly constraining the viability of a variety of cosmological theories for the formation of large-scale structure. While the simplest `vanilla' models of single-field inflation appear sufficient to explain current data \cite{0603449}, forthcoming experiments will test this concordance further. One area in which we can expect significant improvement will be in estimates of non-Gaussianity. Single-field inflation predicts that variations in the CMB should follow simple Gaussian statistics to high accuracy.  On the other hand, more natural inflation scenarios which involve  multiple fields and more complex potentials can, in principle, produce large non-Gaussian signatures that may become measurable in future experiments (see, for example, refs.~\cite{0208055,0511041, multifield,0404084,0312100}). Of course, the CMB power spectrum is not sensitive to these non-Gaussianities so we must look to other higher order correlators.  The bispectrum is the three-point correlator of the multipoles and we note that statistics based on it have been shown to be optimal for testing non-Gaussianity for a variety of inflationary scenarios \cite{0503375,0606001}.

The aim of this paper is to provide a framework for evolving the primordial bispectrum (the bispectrum at the end of inflation) until it leaves an observable imprint in the CMB today. Current theory dictates this can be done by solving a four dimensional integral over the primordial bispectrum multiplied by six highly oscillatory functions.  In practice this is not possible to evaluate numerically.  In section 2 we detail the current most popular approach \cite{0005036} which assumes a simple separable form for the bispectrum, so that the integral separates and becomes tractable.  These methods are good for obtaining quick estimates but their applicability is limited by their ability to fit a suitable analytic approximation for the shape of the primordial bispectrum.  In section 3, we detail a different more general  approach. As the primordial bispectrum is the most important ingredient in the calculation, we place no restrictions on it other than an overall separable scale dependence. From here we reduce the four-dimensional integral to a two-dimensional integral over the primordial bispectrum multiplied by two one-dimensional integrals, one model dependent and one purely geometric.

In section 4, we investigate analytic solutions for the bispectrum. There are two asymptotic categories for the shape function -- ``local'' and ``equilateral'' -- depending on whether the non-Gaussianity is created primarily on superhorizon scales or at horizon-crossing, respectively. In the large angle approximation and for the simplest form of the local bispectrum, there exists an exact solution which can be used to test the accuracy of our numerical methods. While there is no similar solution for the equilateral scenario,  we argue that our methods should be more precise in this case. We have developed code that will accurately evolve the bispectrum from the end of inflation until today, evaluating its angular counterpart in the CMB.  Note that our primary focus here is on the effect of primordial non-Gaussianity on the CMB, so it suffices to use linear radiation transfer functions for this
purpose.   In section 5, we detail the inner workings of the code and point out the key innovations required to compute the integral quickly and accurately.  In section 6, we test against the local analytic solution and demonstrate an accuracy of less than 1\% for multipole values less than $l < 550$ (using the idealized large angle transfer function). We also study the bispectrum for the equilateral case and find it has better convergence with an error of less than  1\% for $l$ values, $l < 1300$. With the correct radiation transfer functions used at small angles, we find that calculations of the full CMB bispectrum are numerically tractable at large $l$ and that the errors remain below 1\% out to multipoles $l<1800$.   We show that growing errors for even larger $l$ are linked directly to the resolution and the asymptotic cut-offs employed.  If the parameters governing these are tightened, arbitrary accuracy can be achieved though at some computational cost.    We provide plots of the CMB bispectrum for the full radiation transfer function for both the local and equilateral cases, showing oscillatory behaviour analogous to the angular power spectrum.  Having established the accuracy and efficiency of these methods, we detail future directions for this work in section 7.

\section{Bispectrum}

We will quickly review the calculation for the bispectrum today from a given primordial bispectrum, this section loosely follows the derivation in \cite{9907431}. The temperature fluctuations in the CMB can be decomposed into spherical harmonics, $Y_{lm}$
\eq
\frac{\Delta T}{T}(\un) = \sum_{l m} a_{lm} Y_{lm}(\un).
\qe
We can invert this expression to find the $a_{lm}$'s using the spherical harmonic orthogonality condition,
\eq
a_{lm} = \int d\un \frac{\D T}{T}(\un) Y^*_{lm}(\un).
\qe
The temperature anisotropy can be decomposed into Legendre polynomials $P_l(\mu)$ in $k$ space,
\eq
\frac{\Delta T}{T}(\un) = \int \frac{d^3k}{(2\pi)^3} \sum^{\infty}_{l=0} (-i)^l (2l+1) \Y (\bk) \D_l(k) P_l(\uk \cdot \un),
\qe
where $\Y (\bk)$ is the primordial gravitational-potential perturbation and $\D_l(k)$ is the radiation transfer function.
We replace the Legendre polynomials using the spherical harmonic addition theorem,
\eq
P_{l}(\uk \cdot \un) = \frac{4 \pi}{2l +1} \sum^{l}_{m=-l} Y_{lm}(\uk) Y^*_{lm}(\un).
\qe
Making these substitutions, and using the orthogonality of the spherical harmonics, we have,
\eq a_{lm}
\nonumber & = & \int \frac{d^3 k}{(2\pi)^3} d^3n \sum_{l'} (-i)^{l'} (2l'+1) \Y(k) \D_{l'}(k) P_{l'}(\u) Y^*_{lm}(\bn)\\
\nonumber & = & \int \frac{d^3 k}{(2\pi)^3} 4\pi \sum_{l'} (-i)^{l'} \Y(k) \D_{l'}(k) \sum_{m'} Y^*_{l'm'}(\uk) \int d^3n Y_{l'm'}(\bn) Y^*_{lm}(\bn)\\
          & = & 4\pi (-i)^l\int \frac{d^3 k}{(2\pi)^3} \Y(k) \D_l(k) Y^*_{lm}(\uk)
\qe
The bispectrum is defined as the three-point correlator of the $a_{lm}$'s,
\eq
B^{m_1 m_2 m_3}_{l_1 l_2 l_3} \= \<a_{l_1 m_1} a_{l_2 m_2} a_{l_3 m_3}\> \\
\nonumber & = & (4 \pi)^3 (-i)^{l_1 + l_2 + l_3} \int \frac{d^3\bk_1}{(2\pi)^3} \frac{d^3\bk_2}{(2\pi)^3} \frac{d^3\bk_3}{(2\pi)^3} \< \Y(\bk_1) \Y(\bk_2) \Y(\bk_3) \> \\
          &   & \D_{l_1}(k_1) \D_{l_2}(k_2) \D_{l_3}(k_3) Y^*_{l_1 m_1}(\uk_1) Y^*_{l_2 m_2}(\uk_2)  Y^*_{l_3 m_3}(\uk_3).
\qe
where $k_1 = |{\bf k}_1|$,  $k_2 = |{\bf k}_2|$ and  $k_3 = |{\bf k}_3|$. 
The primordial bispectrum is defined as
\eq
\< \Y(\bk_1) \Y(\bk_2) \Y(\bk_3) \> = (2\pi)^3 F(k_1,k_2,k_3) \d(\bk_1+\bk_2+\bk_3)\,,
\qe
where $F(k_1,k_2,k_3)$ is the primordial bispectrum shape. We use the integral form of the delta function,
\eq
\d(\bk) = \frac{1}{(2 \pi)^3} \int e^{i \bk \cdot {\bf x}} d^3x,
\qe
and we replace the exponential with spherical harmonics using the Rayleigh expansion, 
\eq
e^{i \bk_1 \cdot {\bf x}} = 4 \pi \sum_l i^l j_l(k_1x) \sum_m Y_{lm}(\uk_1) Y^*_{lm}(\ux)\,,
\qe
where $x = |{\bf x}|$ and the unit vector $\ux = {\bf x}/x$.  
Substituting, then using the orthogonality of the spherical harmonics and the Gaunt integral,
\eq
\curl{G}^{l_1 l_2 l_3}_{m_1 m_2 m_3} = \int{d\Omega \,Y_{l_1 m_1}Y_{l_2 m_2}Y_{l_3 m_3}} =\sqrt{\frac{(2l_1+1)(2l_2+1)(2l_3+1)}{4\pi}} \( \begin{array}{ccc} l_1 & l_2 & l_3 \\ 0 & 0 & 0 \end{array} \) \( \begin{array}{ccc} l_1 & l_2 & l_3 \\ m_1 & m_2 & m_3 \end{array} \),
\qe
where $\( \begin{array}{ccc} l_1 & l_2 & l_3 \\ m_1 & m_2 & m_3 \end{array} \)$ is the Wigner 3j-symbol. We find the bispectrum becomes
\eq
B^{m_1 m_2 m_3}_{l_1 l_2 l_3} = \(\frac{2}{\pi}\)^3 \curl{G}^{l_1 l_2 l_3}_{m_1 m_2 m_3} \int dx dk_1 dk_2 dk_3 (x k_1 k_2 k_3)^2 F(k_1,k_2,k_3) \D_{l_1}(k_1) \D_{l_2}(k_2) \D_{l_3}(k_3) j_{l_1}(k_1x) j_{l_2}(k_2x) j_{l_3}(k_3x)\,.
\qe
It is worth noting that the Wigner 3j-symbol places some restrictions on the $l$ values.  First, no individual $l$ can be greater than the sum of the remaining two (the triangle condition) and, secondly, that the sum of all three must be even. As we assume statistical isotropy it is more common to work with the angle averaged bispectrum,
\eq
B_{l_1 l_2 l_3} = \sum_{m_1 m_2 m_3} \( \begin{array}{ccc} l_1 & l_2 & l_3 \\ m_1 & m_2 & m_3 \end{array} \) B^{m_1 m_2 m_3}_{l_1 l_2 l_3}.
\qe
Observing that,
\eq
\sum_{m_1 m_2 m_3} \( \begin{array}{ccc} l_1 & l_2 & l_3 \\ m_1 & m_2 & m_3 \end{array} \)^2 = 1,
\qe
we have,
\eq
B_{l_1 l_2 l_3} & = &  \(\frac{2}{\pi}\)^3 \sqrt{\frac{(2l_1+1)(2l_2+1)(2l_3+1)}{4\pi}} \( \begin{array}{ccc} l_1 & l_2 & l_3 \\ 0 & 0 & 0 \end{array} \)\\
\nonumber &   & \int dx dk_1 dk_2 dk_3 (x k_1 k_2 k_3)^2 F(k_1,k_2,k_3) \D_{l_1}(k_1) \D_{l_2}(k_2) \D_{l_3}(k_3) j_{l_1}(k_1x) j_{l_2}(k_2x) j_{l_3}(k_3x).
\qe
Also noticing that the Gaunt integral represents a purely geometrical factor which is independent of the primordial bispectrum we define,
\eq
B^{m_1 m_2 m_3}_{l_1 l_2 l_3} = \curl{G}^{l_1 l_2 l_3}_{m_1 m_2 m_3} b_{l_1 l_2 l_3}\,.
\qe
where $b_{l_1 l_2 l_3}$ is the reduced bispectrum,
\eq b_{l_1 l_2 l_3} \label{redbispectrum}
& = &  \(\frac{2}{\pi}\)^3 \int dx dk_1 dk_2 dk_3 (x k_1 k_2 k_3)^2 F(k_1,k_2,k_3) \D_{l_1}(k_1) \D_{l_2}(k_2) \D_{l_3}(k_3) j_{l_1}(k_1x) j_{l_2}(k_2x) j_{l_3}(k_3x).
\qe
This is the quantity that we will be calculating.

To proceed in general we must solve a four-dimensional integral of the shape function multiplied by six highly oscillatory functions, $\D_{l}$ and $j_{l}$, which is not feasible in practice.  Current approaches overcome this by assuming a simple separable form for the shape function. The most common is that developed in ref.~\cite{0005036} where the primary assumption is that the primordial gravitational potential perturbation can be approximated by,
\eq \label{localansatz}
\Y(x) \px \Y_L(x) + f_{NL}\( \Y^2_L(x) - \<\Y^2_L(x)\> \),
\qe
with $\Y_L(x)$ the linear Gaussian part and $f_{NL}$ a constant.  Following this through gives a shape function of 
\eq
F(k_1,k_2,k_3) = 2 f_{NL} \( P^{\Y}(k_1) P^{\Y}(k_2) + P^{\Y}(k_2) P^{\Y}(k_3) + P^{\Y}(k_3) P^{\Y}(k_1)\).
\qe
Substituting this into \rf{redbispectrum} we can now separate the integral into,
\eq \label{decomposition}
\int x^2 dx \; b^L_{l_1}(x) b^L_{l_2}(x) b^{NL}_{l_3}(x) + perms
\qe
with
\eq
b^L_{l}(x) \= \frac{2}{\pi} \int k^2 dk P^{\Y}(k) \D_l(k) j_l(kx)\\
b^{NL}_{l}(x) \= \frac{2}{\pi} f_{NL} \int k^2 dk \D_l(k) j_l(kx)\,,
\qe
that is, it becomes products of  one-dimensional integrals which are manageable numerically.

Searches for the bispectrum in the CMB have consequently focused on establishing limits on $f_{NL}$ via purpose-built statistics rather than calculation of the bispectrum directly.  The WMAP team have constrained non-Gaussianity \cite{0603449} by narrowing their analysis to the local ansatz (\ref{localansatz}), estimating the limits to be $-54 < f_{NL} < 114$.  (This analysis has been repeated in ref.~ \cite{0606001} with an improved estimator to obtain the limits $-36 < f_{NL} < 100$.)   Now primordial non-Gaussianity can be broadly classified into either of two asymptotic regimes, ``local'' or ``equilateral''  \cite{0405356}.  The local case describes models in which the non-Gaussianity is produced outside the Hubble radius (`superhorizon') with examples including the curvaton model \cite{0208055} and multiple field inflation models \cite{0511041}.  The equilateral case describes the case in which the primary contribution to non-Gaussianity arises just as the perturbations leave the 
Hubble radius (`horizon-crossing'), with examples including DBI model \cite{0404084} and ghost inflation \cite{0312100}.  Following the `local' WMAP approach described above, there has also been a
parallel analysis for the equilateral case \cite{0509029}.   The key element shared by this approach is an  analytic approximation for the shape function that allows the separation of the bispectrum integral into the product of three one-dimensional integrals over $k_i$ inside an integral over $x$ as in (\ref{decomposition}).  They then proceed by generating a statistic sensitive to their ansatz to establish limits on a generalisation of the non-Gaussianity parameter $f_{NL}$.  They find considerably weaker constraints on $f_{NL}$ in the equilateral case, $-256 < f_{NL} < 332$ \cite{0606001}, but argue that this represents approximately the same ``level of non-Gaussianity''.    Other recent work in this area includes
ref.~\cite{0509098} where, again using the separable local ansatz (\ref{decomposition}), the particular scale dependence of $f_{NL}$ generated by post-inflationary gravitational evolution is incorporated.   We note also for the local assumption (\ref{localansatz}) in ref.~\cite{0306248}, map-making methods are developed to create individual realisations incorporating the bispectrum and higher order correlators.   

In this paper we have adopted a different more general approach which does not rely on  separable ans\"atze giving rise to the triple decomposition (\ref{decomposition}).  Rather than generating approximations to the bispectrum to allow for easy analytic calculation, instead we have written code that will directly solve the integral for any shape function assuming only the separation of an overall scale dependence. We anticipate that this analysis will encompass all viable inflationary models.

\section{New parametrisation}
\begin{figure}[t]
\centering
\includegraphics[height=2.4in]{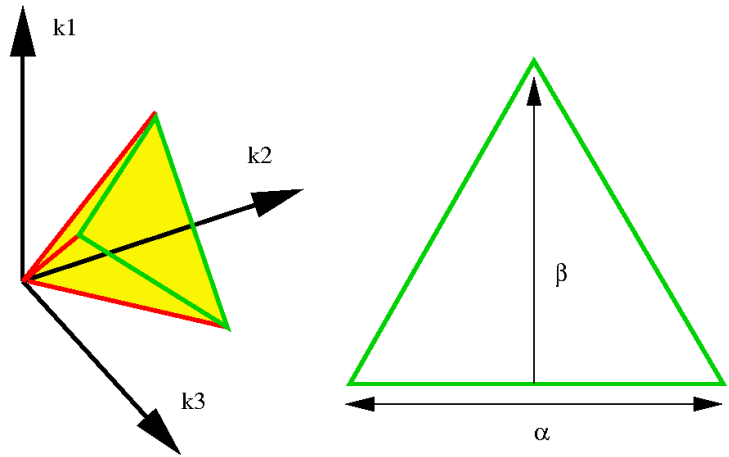}
\caption{\label{triangle}\small Region of integration for the bispectrum calculation. The red lines are $k_1=k_2,\, k_3=0;\; k_2=k_3,\, k_1=0;\; k_3=k_1,\, k_2=0$ and the region of integration is in yellow.  This area can be parametrised into slices represented by the green triangle and the distance $\frac{2}{\sqrt{3}}k$ of the centre of the triangle from the origin.}
\end{figure}

By looking at the expression for the reduced bispectrum (\ref{redbispectrum}), we can rearrange the terms as follows:
\eq \label{biint1}
\(\frac{2}{\pi}\)^3 \int dk_1 dk_2 dk_3 (k_1 k_2 k_3)^2 F(k_1,k_2,k_3) \D_{l_1}(k_1) \D_{l_2}(k_2) \D_{l_3}(k_3) \( \int x^2 dx j_{l_1}(k_1x) j_{l_2}(k_2x) j_{l_3}(k_3x)\).
\qe
The integral in the brackets arises as a direct result of the delta function and analytic solutions have been found, \cite{9107011,9309023,0506114}.  These solutions demonstrate that the three lengths $k_1,k_2,k_3$ must be such as to be able to  form a triangle, as expected from the delta function, which restricts us to the region of $k$-space highlighted in figure~\ref{triangle}.
The analytic solutions also state that the value on the boundary is half that of a point immediately interior. It is natural to reparametrise this region into triangular slices \cite{0410486},
\eq\label{parameters}
\nonumber k_1 = &k a& = k \(1-\b\) \\
\nonumber k_2 = &k b& = \hf k \(1 + \a + \b\)\\
k_3 = &k c& = \hf k \(1 - \a + \b\),
\qe
where $a,\,b,\,c$ are shorthand for the corresponding expressions involving $\a, \,\b$. They are not independent of each other as $a + b + c = 2$. Here, the overall wavenumber scale is given by the semi-perimeter, $k\equiv \frac{1}{2}(k_1+k_2+k_3)$. The surface $k = const$ defines a plane with normal $(1,1,1)$ at a distance $\frac{2}{\sqrt{3}}k$ from the origin and $\a,\,\b$ parametrise our position on that plane. The new parameters have the following domains $0 \le k < \infty$, $0 \le \b \le 1$, and $ -(1-\b)\le \a \le 1 - \b$.  This parametrisation has volume element $dk_1 dk_2 dk_3 = k^2 dk d\a d\b$ (representing a minor improvement over the original triangular parametrisation given in \cite{0410486}). Making these substitutions we have
\eq
\(\frac{2}{\pi}\)^3\int k^2 dk d\a d\b (a\,b\,c)^2 k^6 F(ak,bk,ck) \D_{l_1}(ak) \D_{l_2}(bk) \D_{l_3}(ck) \( \int x^2 dx j_{l_1}(akx) j_{l_2}(bkx) j_{l_3}(ckx)\).
\qe
and, if we rescale $x$ to absorb $k$, we have
\eq
\(\frac{2}{\pi}\)^3\int d\a d\b (a\,b\,c)^2 \(\int \frac{dk}{k} k^6 F(ak,bk,ck) \D_{l_1}(ak) \D_{l_2}(bk) \D_{l_3}(ck) \)\( \int x^2 dx j_{l_1}(ax) j_{l_2}(bx) j_{l_3}(cx)\).
\qe
Given strong observational limits on the scalar tilt we expect the shape function to exhibit near scale-invariance \cite{0603449,0405356}, so we make the substitution
\eq
F(k_1,k_2,k_3) \px \frac{k^{n}}{k^6} F(\a,\b),
\qe
where $n$ is the bispectrum tilt which is expected to be small. This is the only restriction we place on the shape function, up until this point the discussion has been entirely general.  In fact this is slightly over-restrictive and we should note that the method continues to work if we replace $k^n$ with a more general overall scale dependence $f(k)$. Hence, we can rewrite (\ref{biint1}) as
\eq \label{biint2}
\(\frac{2}{\pi}\)^3\int d\a d\b\,  F^{SI}(\a,\b)\, I^T(\a,\b) \, I^G(\a,\b),
\qe
where,
\eq \label{transferint}
I^T(\a,\b) &\equiv& \int \D_{l_1}\(a k\) \D_{l_2}\(b k\) \D_{l_3}\(c k\) k^{n} \frac{dk}{k} \\
I^G(\a,\b) &\equiv& \int j_{l_1}\(a x\) j_{l_2}\(b x\) j_{l_3}\(c x\) x^2 dx \label{geometricint}\\
F^{SI}(\a,\b) &\equiv& \(abc\)^2 F(\a,\b),\label{scaleinvt}
\qe
where we have defined the ``scale independent'' shape function, $F^{SI}$. (A similar concept was discussed in ref.~\cite{0405356}, but can be contrasted with the definition (\ref{scaleinvt}) in our more symmetric triangular parametrisation.) 

The end result  is that we have broken up the four-dimensional integral for the reduced bispectrum  (\ref{redbispectrum}) into a two-dimensional integral (\ref{biint2}), restricted to an equilateral triangle, over the scale-independent shape multiplied by two one-dimensional integrals (\ref{transferint})--(\ref{geometricint}). The first $I^T(\a,\b)$  is over the transfer functions and the second is purely geometric $I^G(\a,\b)$. These two one-dimensional integrands consist entirely of highly oscillatory functions.  However, as we shall discuss later, since they are only one-dimensional it is feasible to evaluate them accurately numerically, freeing us from the severe difficulties previously encountered.

\begin{figure}[t]
\centering
\begin{tabular}{cc}
\includegraphics[height=2.25in]{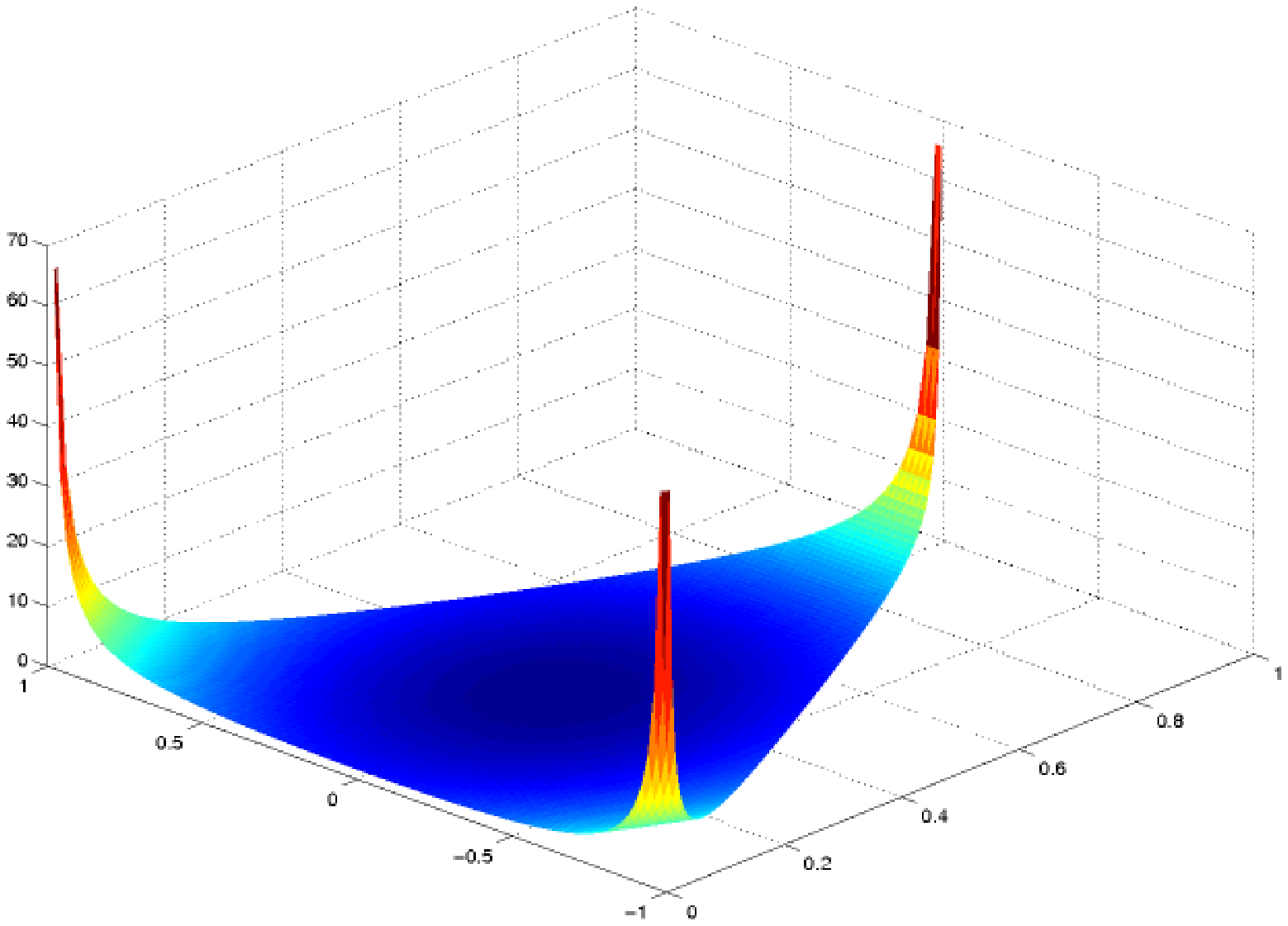} &
\includegraphics[height=2.25in]{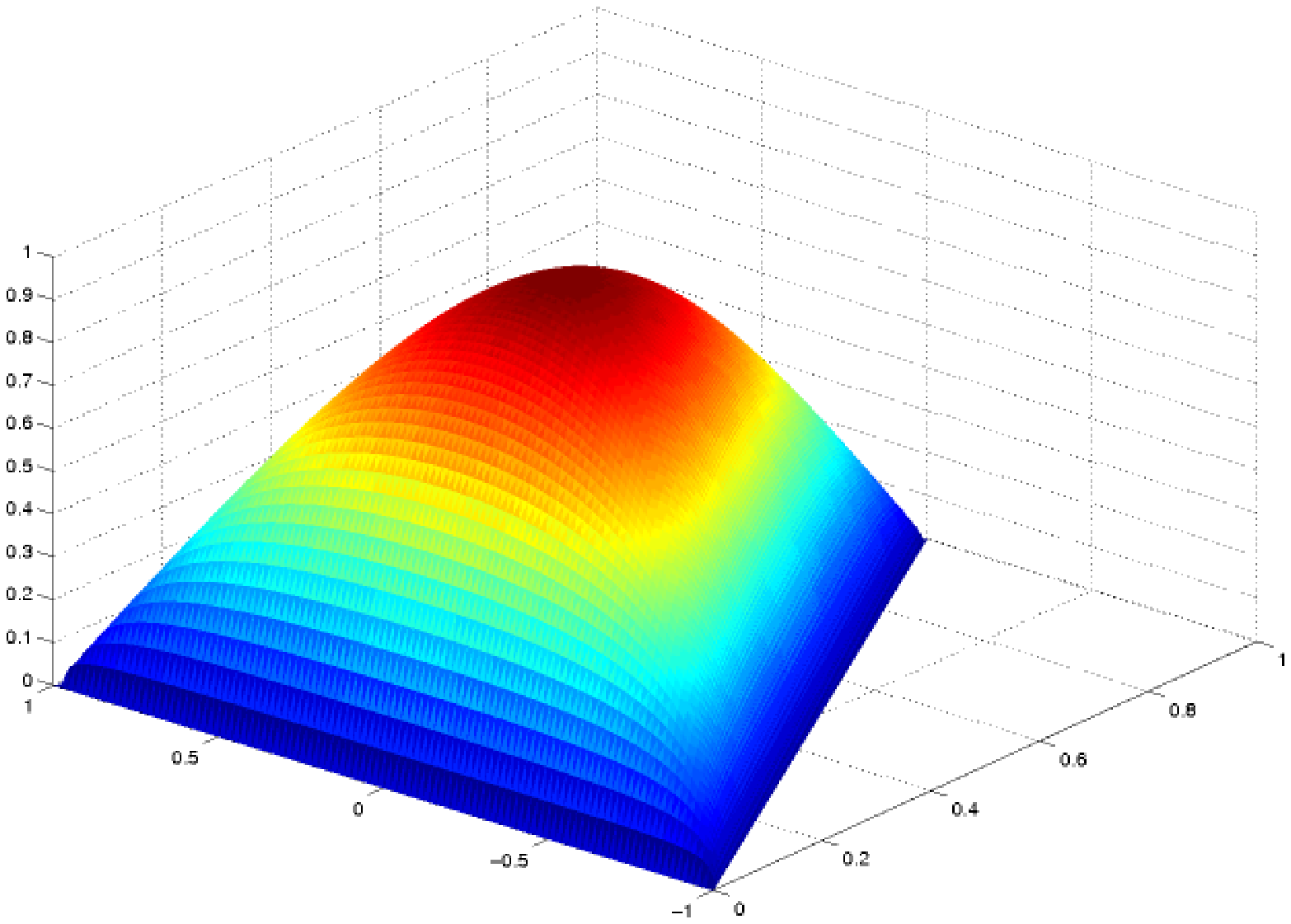} \\
\end{tabular}
\caption{\label{primbisp}\small $F^{SI}$ plotted on the $\a\b$-triangle for the `local' superhorizon shape (left) and the `equilateral' horizon-crossing shape (right). The equilateral case has been scaled to that the centres of both plots are at the same height.}
\end{figure}

The two basic asymptotic limits (local and equilateral) for the scale-independent shape can be approximated by the analytic expressions \cite{0509029}:
\eq\label{ansatze}
F^{SI}_{local}(a, b, c) \= \frac{a^3+b^3+c^3}{a \,b \,c}\\
F^{SI}_{equilateral}(a, b, c) \= \frac{(1-a)(1-b)(1-c)}{a \,b \,c}.
\qe
We have plotted these two bispectrum shapes in figure \ref{primbisp}.   While 
much emphasis has been placed on these two limits, we note  in passing the caveat that 
even single-field inflation yields a more complicated non-Gaussian shape \cite{0210603}.

We will take this opportunity to define a general momentum-dependent $f_{NL}$ that will work in both the local and the equilateral cases (as in \cite{0506704})
\eq
2 f_{NL}(k_1,k_2,k_3) = \frac{F(k_1,k_2,k_3)}{P^{\Y}(k_1)P^{\Y}(k_2) + P^{\Y}(k_2)P^{\Y}(k_3) + P^{\Y}(k_3)P^{\Y}(k_1)}\,.
\qe
This agrees with the definitions given in both \cite{0005036} and \cite{0509029}, given their respective assumptions, but it can now be applied across the full $k_i$ parameter space.   Further, if we make the same parameter substitutions as those used in obtaining (\ref{biint2}), we see that $f_{NL}$ has an overall scale dependence given by $n_{NL} = n +2(1 -  n_s)$ where $n_s$ is the usual scalar spectral index of the power spectrum.

\section{Analytic results}

As well as the  analytic approximations for the shape function in the local and the equilateral limits (\ref{ansatze}), on large angles ($l\ll 200$) the transfer function can be approximated by,
\eq\label{largeangle}
\D_{l}\(k\) = \frac{1}{3} j_{l}\(\D\n \, k\),
\qe
where $\D\n$ is the conformal time period between today $\eta_0$  and the surface of last scattering
$\eta_{\rm dec}$. Substituting these into the bispectrum,  the whole integrand can be expressed in 
closed form and so we can look for analytic solutions.  In the local case, we make the replacement
\eq
F^{SI}(k_1,k_2,k_3) = \frac{k_1^2}{k_2 k_3} + \frac{k_2^2}{k_3 k_1} + \frac{k_3^2}{k_1 k_2}.
\qe
and the bispectrum integral separates into three integrals over the $k_i$'s inside an integral over $x$,
\eq
\(\frac{2}{3\pi}\)^3 \int x^2 dx \(\int k_1^2 dk_1 j_{l_1}(\D \n k_1)j_{l_1} (x k_1)\) \(\int \frac{dk_2}{k_2}j_{l_2}(\D \n k_2)j_{l_2}(x k_2)\) \(\int \frac{dk_3}{k_3}j_{l_3}(\D \n k_3)j_{l_3}(x k_3)\),
\qe
plus permutations. The $k_1$ integral evaluates to $\frac{\pi}{2 x^2} \d(x-\D\n)$ and integrating out the delta function leaves,
\eq
\(\frac{4}{27\pi^2}\) \(\int \frac{dk_2}{k_2}j^2_{l_2}(\D \n k_2)\) \(\int \frac{dk_3}{k_3}j^2_{l_3}(\D \n k_3)\).
\qe
These two integrals evaluate to $(2l(l+1))^{-1}$ so with permutations we have the analytic result
\eq \label{analy}
\(\frac{1}{27\pi^2}\) \(\frac{1}{l_1(l_1+1)l_2(l_2+1)} + \frac{1}{l_2(l_2+1)l_3(l_3+1)} +  \frac{1}{l_3(l_3+1)l_1(l_1+1)} \)
\qe
This exact solution (in the large angle approximation) provides an excellent reference for checking the accuracy of our numerical methods in later sections.

For the equilateral case, we make the replacement
\eq
F^{SI}(k_1,k_2,k_3) \= \frac{1}{8} \frac{(-k_1+k_2+k_3)(k_1-k_2+k_3)(k_1+k_2-k_3)}{k_1 k_2 k_3} \\
&=& \frac{1}{8} \(-2 - \frac{k_1^2}{k_2 k_3} + \frac{k_1}{k_2} + \mbox{all possible permutations}\).
\qe
When we substitute this into the full integral it again separates into three integrals over the $k_i$'s inside an integral over $x$.  The middle term is the same as the local case with the exact solution given above.  For the remaining two terms we are left to solve integrals like,
\eq
\int dk k^n j_{l}(k)j_{l}(x k) = \frac{\pi}{2\sqrt{x}}\int dk k^{n-1} J_{l+\hf}(k) J_{l+\hf}(x k),
\qe
for $n=1,\,0,\,-1$.  The case $n=0$ has a nice solution \cite{6601405} (p405),
\eq
x>1 \imp \frac{\pi}{2}\frac{x^{-(l+1)}}{2l+1} &:& x<1 \imp \frac{\pi}{2}\frac{x^{l}}{2l+1}.
\qe
Unfortunately, however, both $n=1,\,-1$ produce complicated expressions involving 
hypergeometric functions \cite{6601405} (p401),
\eq
n=1 &:& x<1 \imp \frac{\sqrt{\pi}\,\G\(l+1\)}{2\,\G\(l+\frac{3}{2}\)} x^l \;{}_2F_1\(\hf,l+1;l+\frac{3}{2};x^2\) \\
&& x>1 \imp \frac{\sqrt{\pi}\,\G\(l+1\)}{2\,\G\(l+\frac{3}{2}\)} x^{-(l+2)} \;{}_2F_1\(\hf,l+1;l+\frac{3}{2};\frac{1}{x^2}\)\\
n=-1 &:& x<1 \imp \frac{\sqrt{\pi}\,\G\(l\)}{4\,\G\(l+\frac{3}{2}\)} x^l \;{}_2F_1\(-\hf,l;l+\frac{3}{2};x^2\) \\
&& x>1 \imp \frac{\sqrt{\pi}\,\G\(l\)}{4\,\G\(l+\frac{3}{2}\)} x^{-l} \;{}_2F_1\(-\hf,l;l+\frac{3}{2};\frac{1}{x^2}\).
\qe
so apparently there is no analogous route to a simple analytic result.

We have also pursued an alternative approach by deriving series solutions for the integrals required for the 
bispectrum.  These are presented in the Appendix, along with a minor correction to the literature, but 
their utility is yet to be established since they exhibit slow numerical convergence in key limits. 

\section{Numerical solutions}

The main problem with a numerical evaluation of the bispectrum is the intensive sampling required to deal with the extreme oscillatory nature of the integrand.  Recall the progress made so far with a 4D integral over six highly oscillatory functions broken down into two 1D integrals (\ref{transferint}) and (\ref{geometricint}) over three highly oscillatory functions each, inside a 2D integral (\ref{biint2}) over a fixed triangular domain.  This is a significant improvement because it has reduced the overall dimensionality of the problem and limited the worst oscillatory behaviour to one dimension. However, we still have to work hard to get the integrals to converge fast enough for the calculation to be feasible. At each point in the $\a\b$-triangle we must solve the two 1D integrals, $I^T(\a,\b)$ and $I^G(\a,\b)$ with each integrand composed of either Bessel or transfer functions which are non-trivial to calculate. Fortunately, the only part of the integrand that changes as we move about the $\a\b$-triangle (i.e.\ at fixed $l_1,\, l_2,\,l_3$) are the values of $a,\,b,\,c$ (see (\ref{parameters})).  These simply rescale the three functions relative to each other so we do not need to re-evaluate them at each point.  Instead we can calculate the functions once beforehand, retaining them for future reference. At each point in the $\a\b$-triangle, we then rescale and multiply them together to form the integrand, saving lengthy calculation.

\begin{figure}[t]
\centering
\includegraphics[height=2.75in]{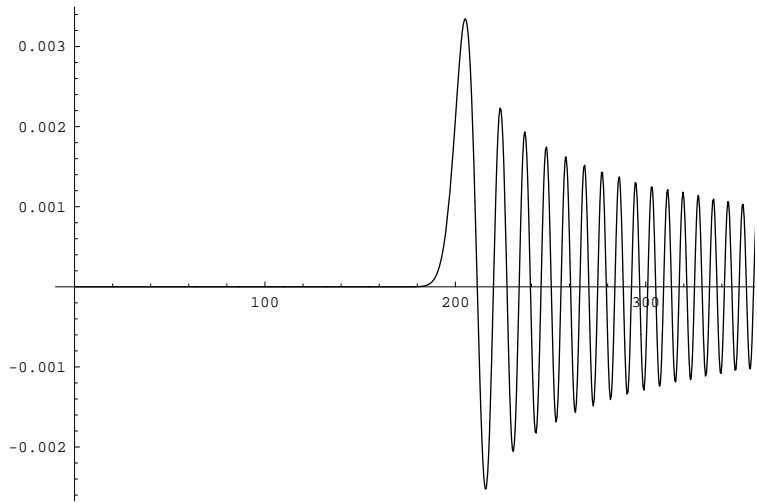}
\caption{\label{sb200}\small A plot of the spherical Bessel function $l=200$.  Note how the function essentially vanishes up to $l\approx 180$.}
\end{figure}

The preliminary calculations of the Bessel and transfer functions are performed for a closely spaced range of $l$-values with the results stored in a table, which can then be read in prior to every bispectrum calculation.  The transfer functions for all $l$'s were calculated using the line-of-sight code CMB2000 (or else an equivalent such as CMBFAST or CAMB), along with the associated spherical Bessel function at the same $l$, and stored in 
separate tables.  Next we take the tabulated functions and interpolate using a cubic spline to create a dense grid of values.  Bessel and transfer functions for large $l$ start with a long flat section which is approximately zero before the first peak and hence does not contribute to the integral \rfig{sb200}. This section is set identically to zero with a view to skipping it in the later integration; we define this as the part for which the function magnitude remains $10^{-10}$ times below its maximum value.

Now consider the two-dimensional integral (\ref{biint2}) over the $\a\b$-equilateral triangle at fixed
multipoles $l_1,\, l_2,\,l_3$.   
Having produced the two tables of Bessel and transfer functions we access the relevant rows for the given multipole set.  At each point in the $\a\b$-triangle we then calculate $a,b,c$ and stretch the data appropriately using linear interpolation to calculate the new points. We skip the first section where the values are zero and then calculate the two one-dimensional integrals using the trapezoidal rule.  This stretching and linear interpolation will tend to magnify any errors in our original data, so 
this is mitigated by using a very large number of points (${\cal O}(10^6)$), even though the integrands use 
a much smaller selection (say every 20th point).  This balances accuracy, using dense initial sampling, with speed, using sparser sampling for the integration. Next we recall that for the geometric integral the value on the boundary is half that of the interior, causing slow convergence of the two-dimensional integral near the discontinuous edges.   We avoid this problem by doubling the result of the geometric integration for points actually on the boundary.  The important further check that needs to be made is when to terminate the integration.  There is a hard boundary when the tabulated data runs out (set by an overall range), but most integrals converge well before this limit.  Convergence is determined by comparing the contribution of the last $x$ sections (typically ${\cal O}(10,000)$) to the value of the total integral.  If it remains unchanged up to a certain percentage tolerance (typically $\px 0.0001\%$) then the integration is stopped.  However, given the  oscillatory nature of the integrand it is important to cover a large region over many periods in order to correctly determine this threshold.  This is most important in the corners of the triangle where the data is stretched, 
along with the period of oscillation, by the largest amount.   As we shall note in the next section,  these 
truncation errors in the 1D integrals can be significant at large $l$ unless handled carefully. 

\begin{figure}[t]
\centering
\includegraphics[height=2.5in]{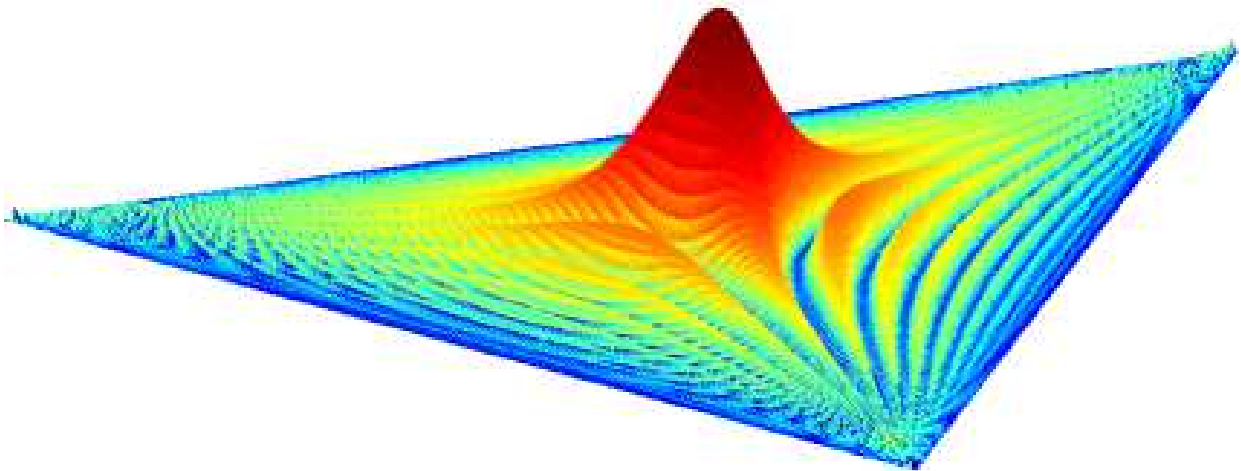}
\caption{\label{202020}\small The equilateral triangle we need to integrate over to obtain the bispectrum for $l_1=l_2=l_3=20$.  Note the small scale structure to the right of the peak.  The first few oscillations make a significant contribution and must be sampled intensively for an accurate result.  The rest of the oscillations, however, do not because of cancellations.  Hence any adaptive algorithm must be able to determine the importance of any structure it finds.}
\end{figure}

\begin{figure}[t]
\centering
\includegraphics[height=1.25in]{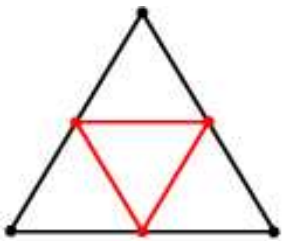}
\caption{\label{refinement}\small Refinement method. We start with the black cell defined by the three black points at the corners.  The cell is then divided into four by calculating the three red points and the change in area is then $\frac{1}{4}\left|\sum black - \sum red\right|$.}
\end{figure}

\begin{figure}[b]
\centering
\begin{tabular}{ccccc}
\includegraphics[height=1.1in]{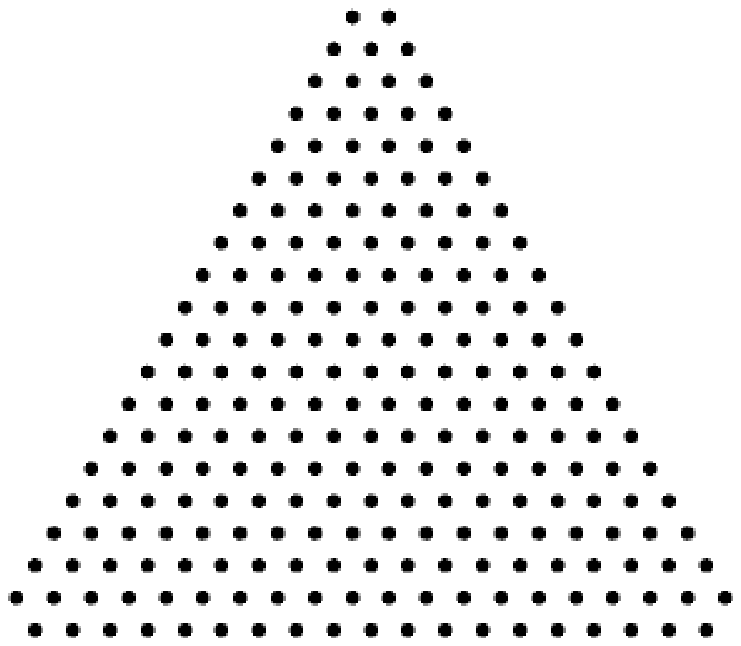}  &
\includegraphics[height=1.1in]{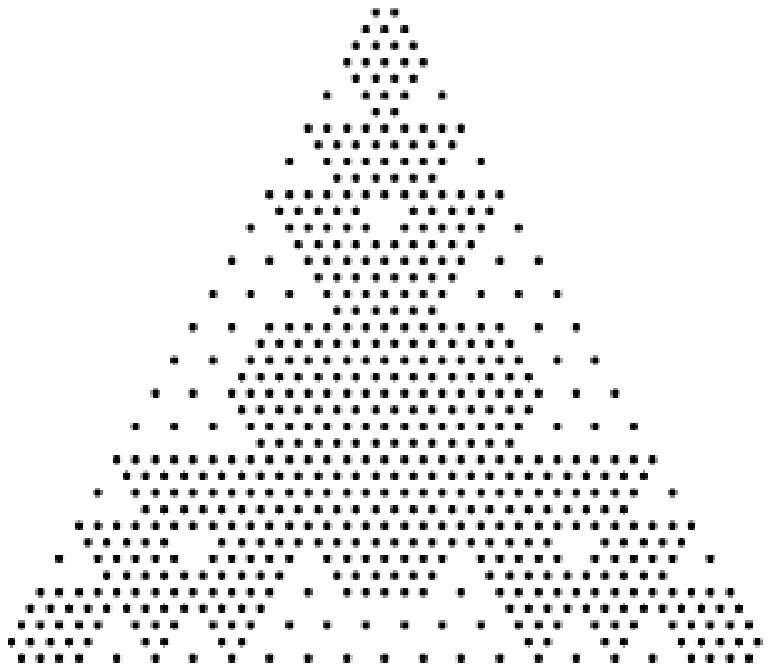}  &
\includegraphics[height=1.1in]{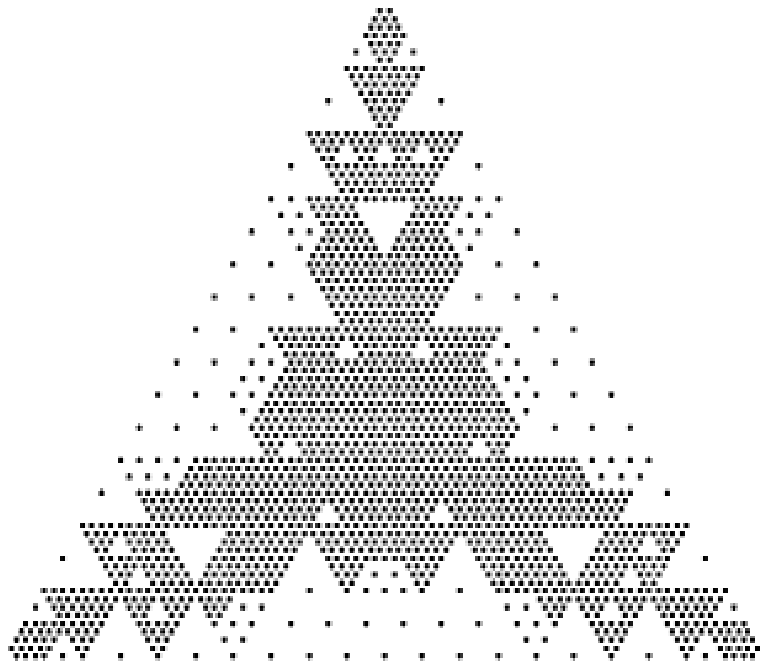}  &
\includegraphics[height=1.1in]{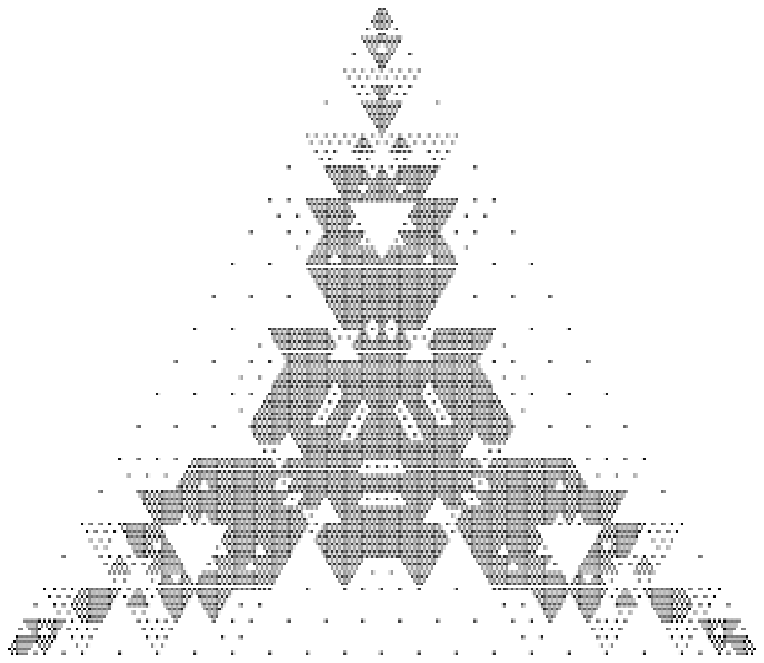}  &
\includegraphics[height=1.1in]{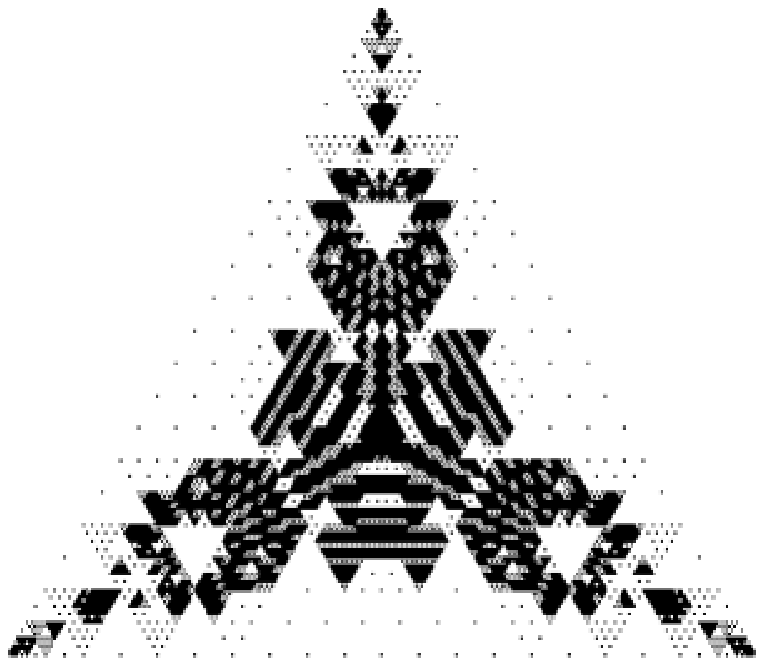}  \\
\end{tabular}
\caption{\label{grid}\small Successive refinement of the grid for integration of the $l_1=l_2=l_3=20$ triangle, the left is a recursion depth of one proceeding to five on the far right. After a recursion depth of five there is little change in the grid or result, however the error for the integration is still 4.45\%.  If we look at the diagram we see a blank triangle about halfway between the corner and the centre.  This region contains important structure but it has been missed as the initial sampling grid was too sparse.  By doubling the initial grid the error is reduced to 0.8\%.}
\end{figure}

When attempting to perform the 2D integral (\ref{biint2}) over the triangle we have to balance two key considerations. Every point  used entails the calculation of both one-dimensional integrals so we want to minimise their number to reduce calculation time.  Unfortunately, while the majority of the highly oscillatory behaviour is confined to the one-dimensional integrals there is still some oscillatory behaviour which carries over to the $\a\b$-triangle \rfig{202020}. As the $\a\b$-triangle has small scale structure we need to use a fine mesh to gain accuracy.  To achieve this we have developed an adaptive method that first uses a sparse triangular grid to estimate the integral using a 2D-trapezoidal rule.  This then steps through each triangular cell and divides it into four, calculating the change in the local integrand contribution \rfig{refinement}.  
If this change is below a tunable accuracy parameter then it uses the original result,  but if it is above then it continues the refinement process by dividing each of these triangles into four new triangles and estimating their contribution. This will be repeated until the change in the local result is below an acceptable limit (or the recursion hits a predefined tunable recursion depth).  The net effect is that we only have dense grids in areas where the most significant contributions to the integral are made, with the remaining areas only sampled sparsely. For the $l_1=l_2=l_3=20$ triangle integration, we achieve convergence for a recursion depth of five after which there is little improvement.  One must be careful, however, to set the original sampling grid to broadly capture most of the structure or else the refinement algorithm may cease prematurely \rfig{grid}.

We note that these calculations  naturally coarse-grain the computational work 
either through the sampling of 1D integrals on the 2D triangular grid or, at a higher level, simply 
by evaluating the bispectrum at different multipole values.   The problem is well-suited to parallelisation 
on a large supercomputer or cluster and this has been achieved with the present code using an MPI implementation which dramatically reduces calculation timescales.

\section{Results}

As we previously noted, bispectrum multipole values $l_1,\,l_2,\,l_3$ are restricted by the same triangle condition as the wavenumbers $k_1,\,k_2,\, k_3$.  This allows us to propose a similar parametrisation to that we used for the $k_i$'s in (\ref{parameters}), that is,
\eq
\nonumber l_1 \= \frac{3}{2} l \(1-\g\) \\
\nonumber l_2 \= \frac{3}{4} l \(1 + \d + \g\)\\
l_3 \= \frac{3}{4} l \(1 - \d + \g\).
\qe
We will use these parameters  to graph the bispectrum in two ways, (i) the equal multipole case $l_1=l_2=l_3$, i.e. the distance from the origin plotting against $l$ (holding $\d=0$ and $\g=\frac{1}{3}$) and (ii) transverse triangular slices with $l=const$.

\begin{figure}[b]
\centering
\includegraphics[height=2.4in]{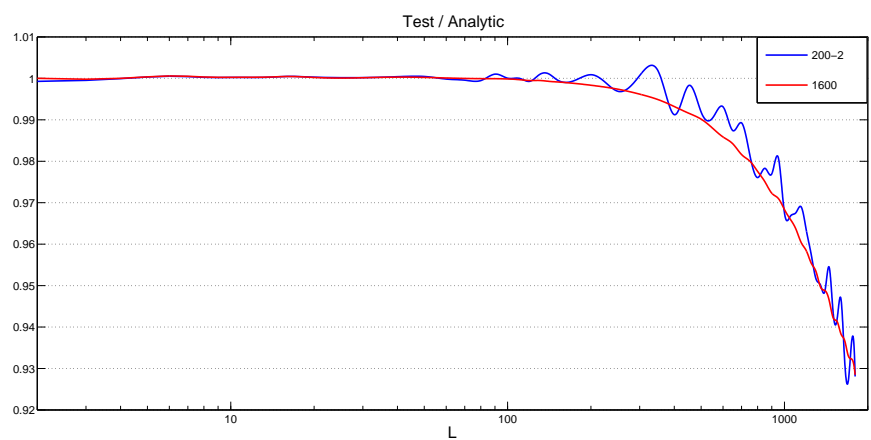}
\caption{\label{test}\small Plot of the bispectrum for the local case in the large angle approximation divided by the analytic result.  The blue line is with the adaptive code starting with a grid of 200 and allowing two recursions, the red is for a grid of 1600.  For the fine 1600 grid the oscillations disappear and the line is smooth indicating that this eliminates error from the grid sampling.  Note the adaptive method remains within a percent of this line.  Both the results are within a percent up to $l\px500$ then we notice a downward trend due to the premature truncation of the one dimensional geometric integral}
\end{figure}

In order to characterise the main sources of systematic error, we begin by comparing to the 
exact solution in the local case (\ref{analy}) using the large angle transfer functions (\ref{largeangle}).
This is a stringent test because the fall-off in the radiation transfer functions is much faster at 
large $l$ in the realistic case; the errors in the test case are more pronounced and help delineate both resolution and truncation errors.  
We have calculated the equal momentum bispectrum from $l=2$ to $l=1800$ on a fairly sparse initial triangular grid of sidelength $N= 200$ and a maximum recursive depth of only 2, and we find that the numerical result has an error less than 1\% for $l < 550$ and a maximum error of 9\% for $l=1800$.  We also repeated the same calculations with a much finer grid with $N= 1600$ to find comparable results 
 which are shown in figure~\ref{test}, demonstrating little improvement from the increased resolution. 
The calculations were performed on the Cosmos supercomputer and took between 16 minutes and 4 hours 55 mins cpu time with a median of  about 80 mins for the adaptive method.  The average 
time to calculate each bispectrum point $b_{lll}$ on the finer grid was greater than 11 hours. 

\begin{figure}[t]
\centering
\includegraphics[height=2.4in]{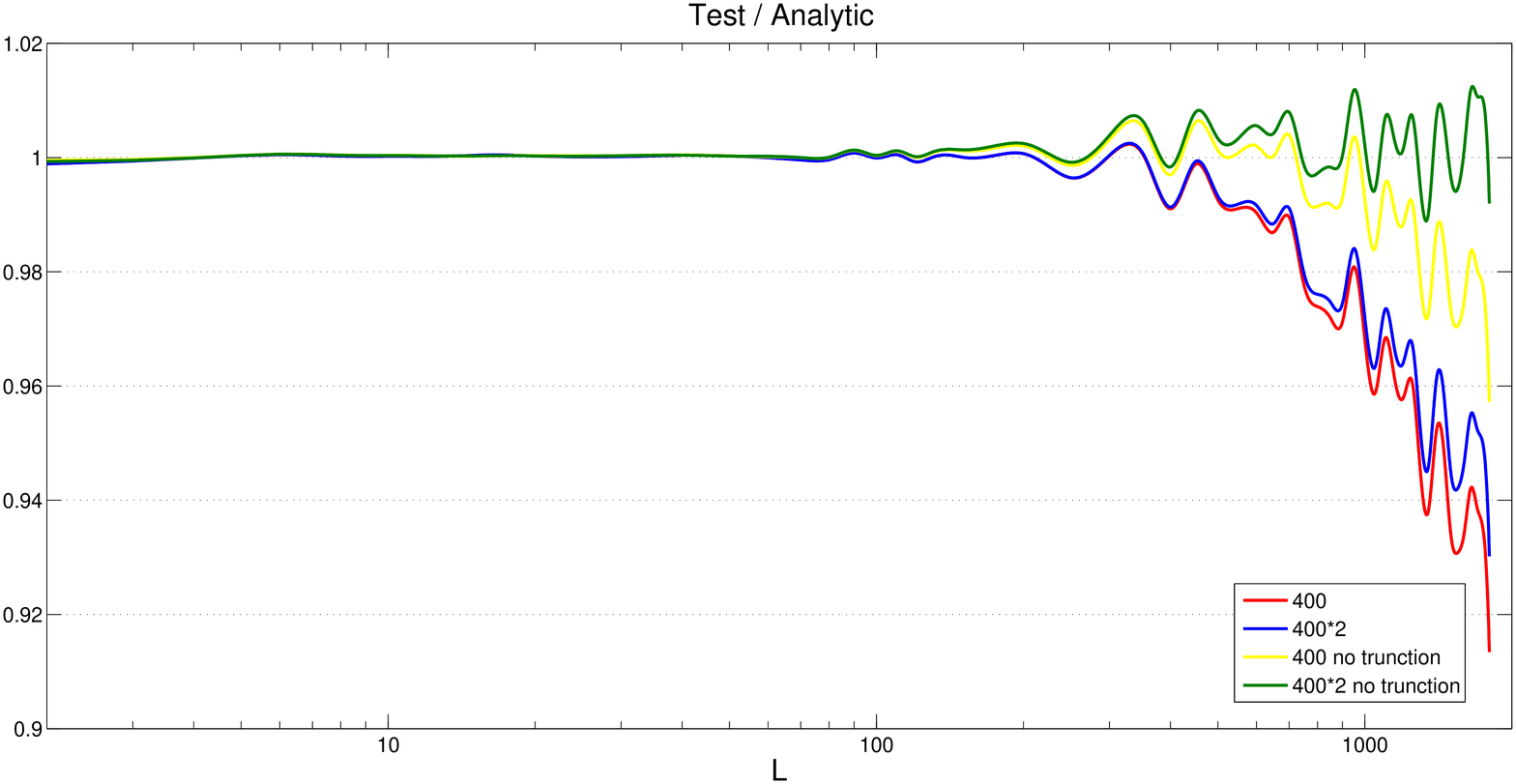}
\caption{\label{test400}\small Sources of error from the one dimensional integrals.  The four lines are all for the same grid of 400 on the triangle.  The red line is the calculation for the standard conditions,  the blue is when we double the region of integration, the yellow is with the intelligent truncation turned off and the green is when we do both.}
\end{figure}

The largest source of error in the calculation is a tendency to underestimate the integral due to premature truncation of the one-dimensional integrals for large $l$.  There are two reasons for this.  First, in the case of $l=1800$ the Bessel function is approximately zero until about 1700.  Truncating at a value of 10000 we only need to rescale by $a < 0.17$ for the non-zero section of the integrand to be shifted off the interval, making the whole integral vanish numerically.  However, this only affects the outside of the $\a\b$-triangle so it only generates a small error.  The second reason is that the intelligent truncation routine that decides if the one-dimensional integral has converged can be too aggressive.   This affects all points and is the primary source
of numerical error.  If we double the integration cut-off and eliminate the intelligent truncation the error in the one dimensional integrals is drastically reduced \rfig{test400}.  The cost is a fivefold increase in calculation times for an adaptive $N= 400$ grid.
This is mainly a problem for the geometric integral as this only decays as $1/x$ requiring late truncation for accuracy.  Here, the large-angle  transfer integrand decays as $1/x^4$ so it converges more quickly.  If we calculate the test case again with an $N=1600$ grid, doubling the 1D region of integration, and turning off the truncation routine, then the error is reduced comfortably within 1\%.  Unfortunately the integrals then take just under 80 hours to complete.

\begin{figure}[t]
\centering
\includegraphics[height=2.4in]{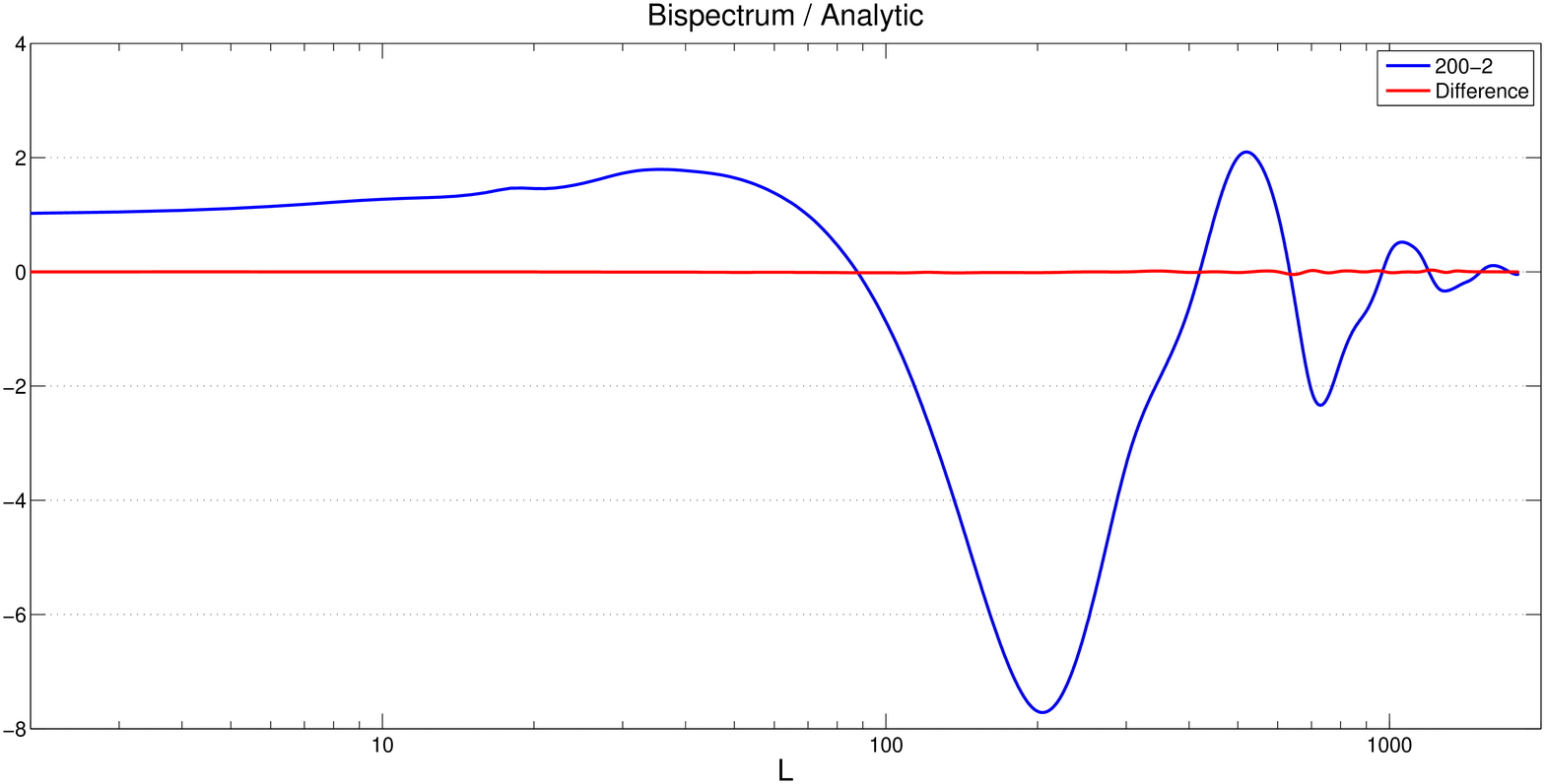}
\caption{\label{bispectrum}\small Plot of the bispectrum in the local case using the full radiation transfer function over the result for the large angle approximation. Note the three main peaks where we would expect from the plot of the power spectrum. The red line is the difference from using the dense grid.  At 650 it reaches a maximum of only 0.045 which justifies our use of the much faster adaptive method.}
\end{figure}

Having confirmed the accuracy of the method, in principle, the calculations were then repeated with the full radiation transfer function.  Here, with an initial $N=200$ grid and a recursion depth of 2, the calculations completed on Cosmos in a median time of 31 mins.  The results for the equal momentum bispectrum are shown in figure~\ref{bispectrum}.   Its behaviour is as expected with the two main peaks appearing at $l\px200$ and $l\px500$, mirroring those observed in the CMB power spectrum.  We have also calculated the bispectrum using a dense $N=1600$ grid with the difference also plotted in figure~\ref{bispectrum}, confirming the absence of a significant resolution error.   On the other hand, we might be concerned about the important truncation error we observed in the large-angle test case discussed above.  To check this we ran long calculations for the $N=1600$ grid but without intelligent truncation and while also doubling the cut-off in the geometric one-dimensional integral.  The results gave a maximum correction of only 0.004\%, indicating that truncation error is also insignificant in the realistic case. 
The explanation for this improvement in accuracy lies in the much more rapid exponential fall-off at large $l$ of the  true radiation transfer function, when compared to the large angle solution (\ref{largeangle}).  Since the transfer function integral multiplies the geometric integral in the final 2D integrand (\ref{biint2}), it also suppresses any errors due to the slow convergence of the latter.  The comparison in figure~\ref{bispectrum} indicates that even the $N=200$ adaptive grid achieves a bispectrum accuracy better than 1\% across the full multipole range $l< 1800$.  

\begin{figure}[t]
\centering
\includegraphics[height=2.4in]{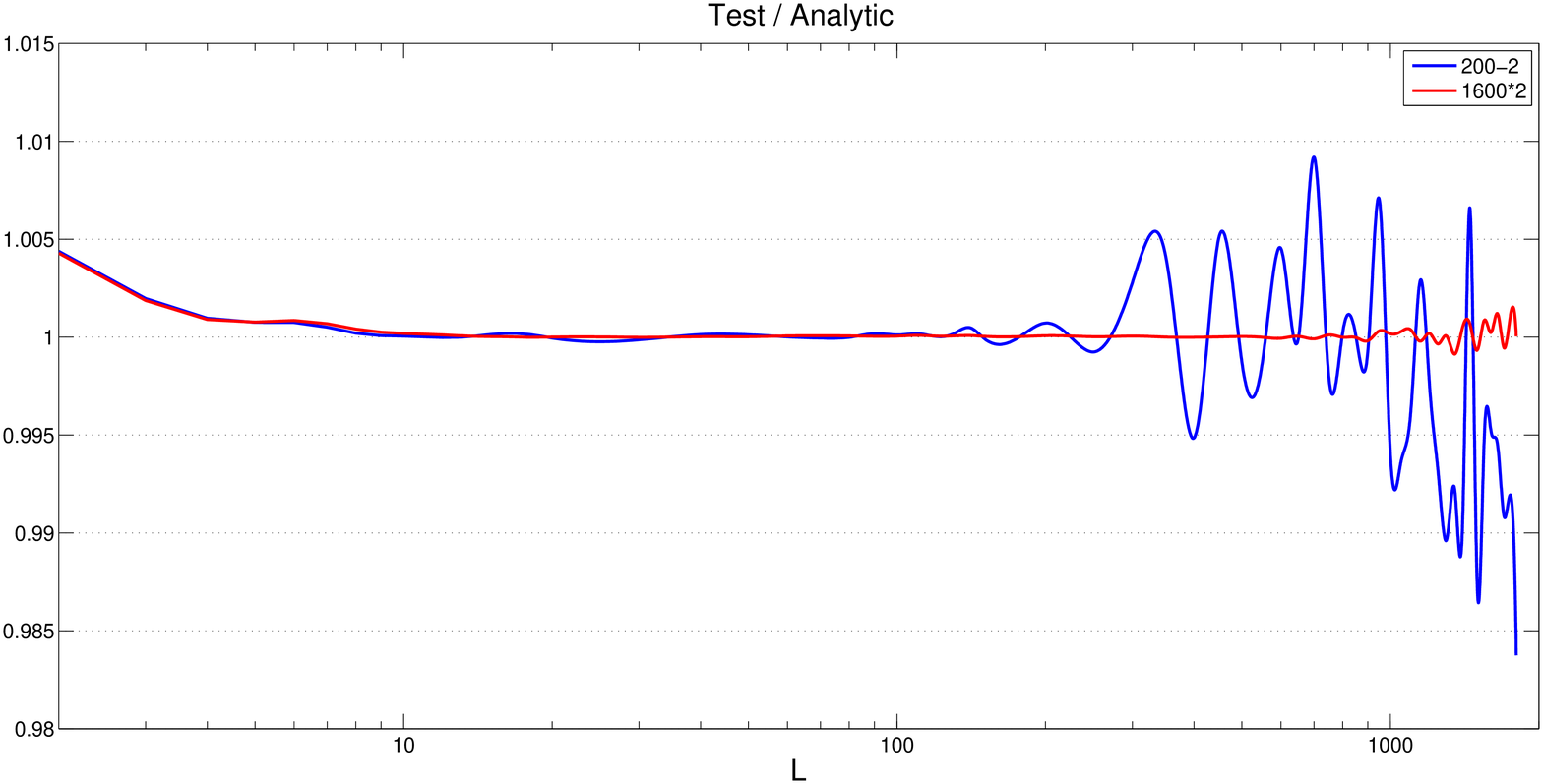}
\caption{\label{test_eq}\small Plot of the bispectrum for the equilateral case in the large angle approximation over the analytic result.  The result has been scaled so that the result for $l=20$ is 1 as this is the most stable part of the plot.  As we have no analytic solution to compare to in the equilateral case we have calculated the bispectrum on a dense $N=1600$ grid without the intelligent truncation and doubling the cut-off.  By comparing the two plots we see the error in within 1\% up until $l=1300$, a significant improvement over the previous case.}
\end{figure}

In the equilateral case we no longer have an analytic solution with which to compare.  However, both $I^T(\a,\b)$ and $I^G(\a,\b)$ are strongly peaked in the centre of the $\a\b$-triangle (as we can see from figure \ref{202020}). Since both bispectrum shapes are flat in the centre, we expect the equilateral case to exhibit approximately the same scaling with $l$ as the local case for $b_{lll}$. To test accuracy we can again take the large angle approximation, dividing by the local analytic solution to normalise.  As anticipated, we find that numerical accuracy is improved over the local test case, because the more oscillatory contributions near the boundaries are 
further suppressed by the primordial shape.  Using an adaptive $N=200$ grid (recursive depth 2), 
we find that this test case has an error of less than 1\% for $l < 1300$ and maximum error of 1.7\% for $l=1800$ when compared to the calculation on a $N=1600$ grid without the intelligent truncation and doubling the cut-off \rfig{test_eq}.

\begin{figure}[t]
\centering
\includegraphics[height=3.0in]{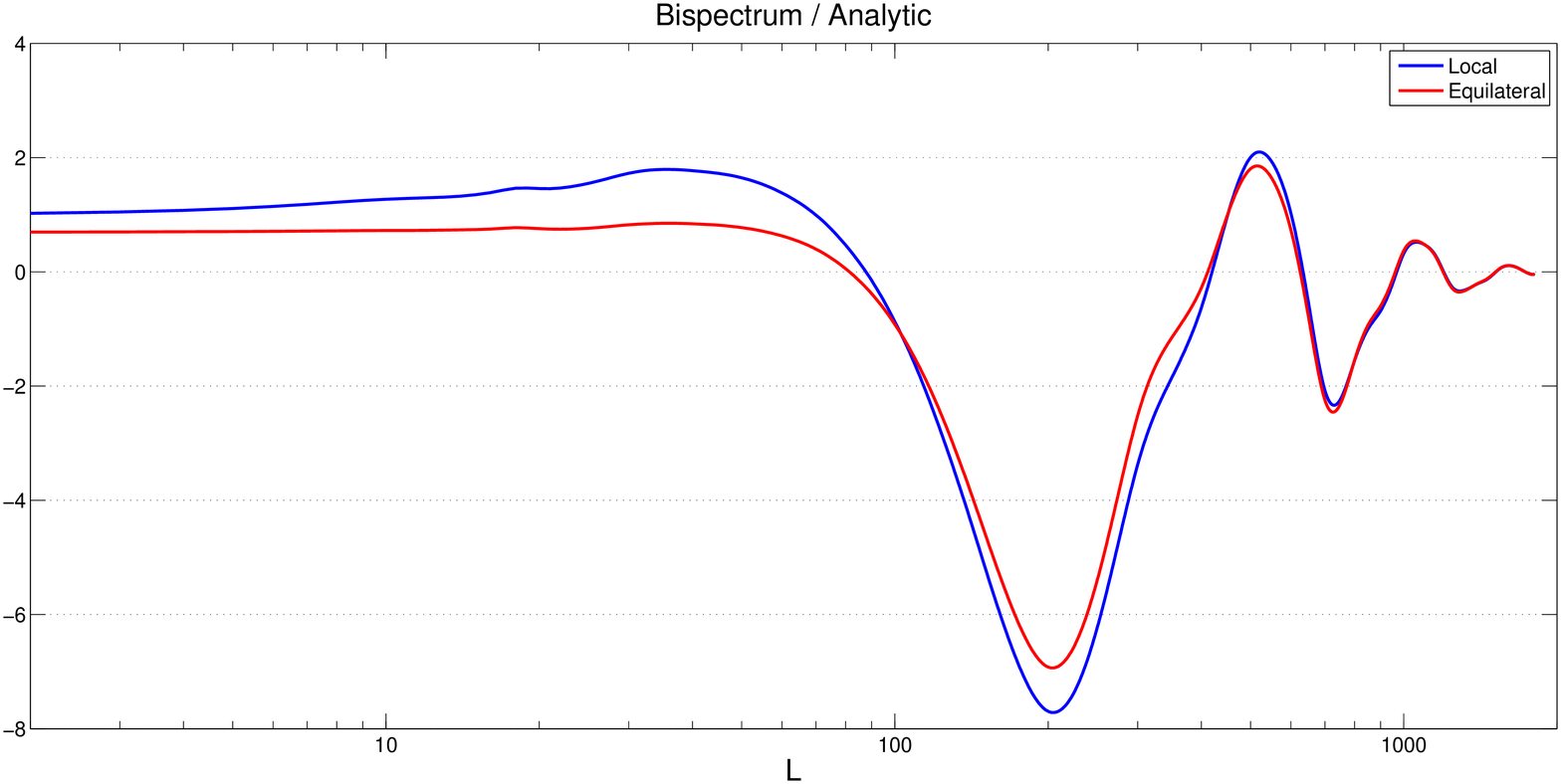}
\caption{\label{bispectrum_eq}\small Plot of the bispectrum for the both the local case and the equilateral case using the full radiation transfer function over the analytic result.  The equilateral case has been multiplied by 24, the ratio between the height of the centre point of the local case to the equilateral case.  We again see the main peaks where we would expect from the power spectrum calculation but the local case is smaller for low $l$.  For large $l$ the two one-dimensional integrals act like delta functions picking out the point on the shape functions where $k_1,\,k_2,\,k_3$ have the same ratio as $l_1,\,l_2,\,l_3$.  As we have scaled the centre points to be the same height the bispectrum curves merge for large $l$.}
\end{figure}

The calculation for the equilateral case with the full radiation transfer function gives much the same results as for the local case \rfig{bispectrum_eq}.  Here we again divide by the analytic solution scaling the result by the factor 24 so that the centres of the two shape functions are both the same height. The main difference between the two is that the peaks are slightly smaller in the equilateral case for low $l$ (figure~\ref{bispectrum_eq}).  For large $l$,  the results are almost identical because the larger the $l$-values the more peaked $I^T$ and $I^G$ become when plotted on the $\a\b$-triangle.  For large $l$ they act like a delta function picking out only the shape function value where $k_1,\,k_2,\,k_3$ have the same ratio as $l_1,\,l_2,\,l_3$.  So for large $l$, the equal $l$ bispectrum is essentially only proportional to the height in the centre of the shape function rather than its shape. These bispectrum calculations on Cosmos took an average of approximately  30 mins for each $l$.

It is clear from figures \ref{bispectrum} and \ref{bispectrum_eq} that it would be difficult to distinguish between the local and equilateral cases on the basis of the equal momentum bispectrum.  Instead, to achieve  this we must look at the transverse triangular slices. For both cases three slices were made at $3l=850$, $3l=1850$ and at $3l=3650$ with points being calculated at $l_i$'s that are multiples of 50.  For the slowly converging test case with a local shape function and the large angle approximation, we find that the error in the calculation for $3l=850$ is less than 0.6\%, for $3l=1850$ less than 3\%, and for $3l=3650$ less than 5\%.  However the error is always negative and its range is small in all cases,  indicating that it is primarily systematic and due to the premature truncation issue discussed above.  Most importantly we note that the error appears to be a function only of the sum, and not the specific combination, of the $l_i$'s so there are no new issues in dealing with the slices.  

\begin{figure}[t]
\centering
\begin{tabular}{cc}
\includegraphics[height=2.3in]{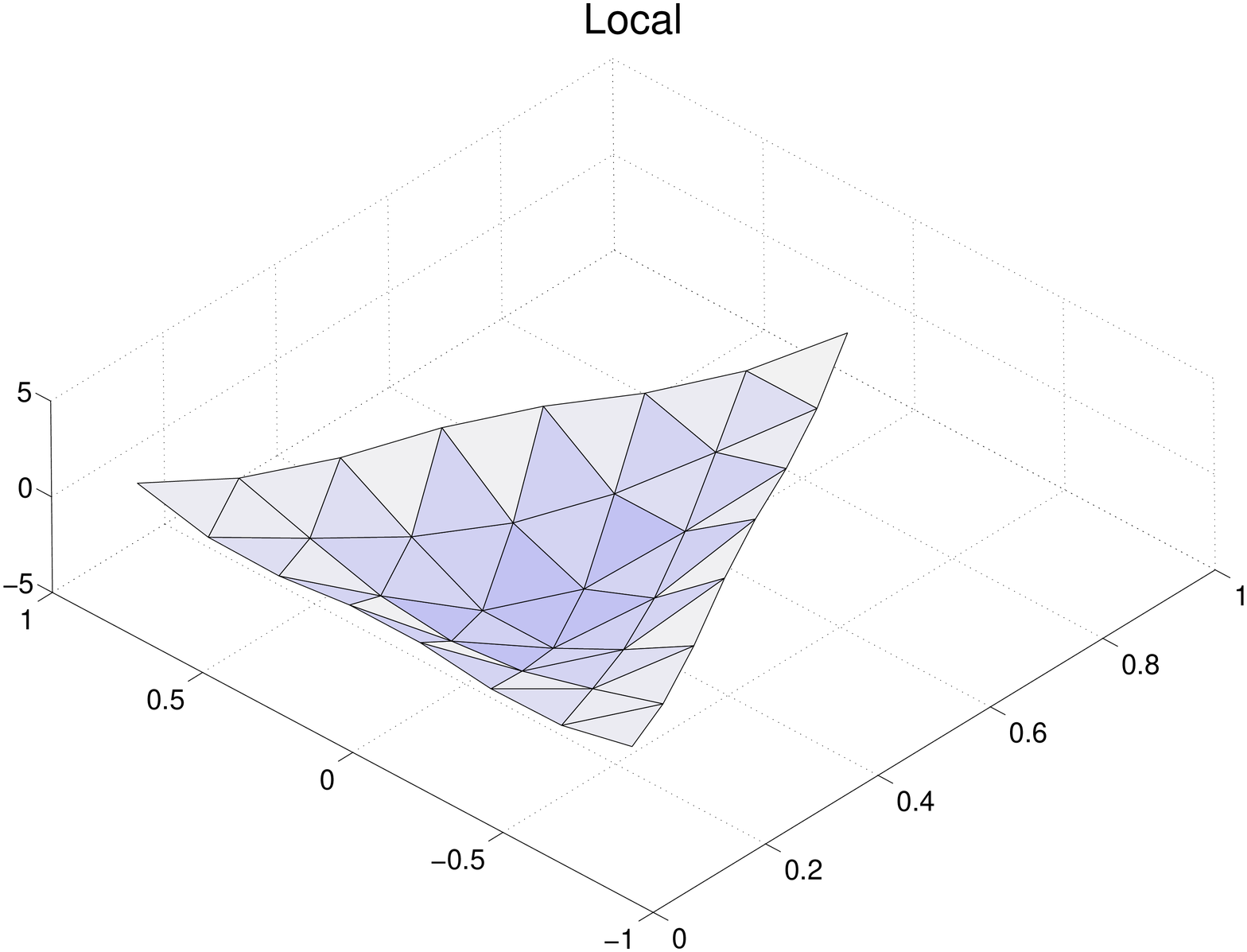} &
\includegraphics[height=2.3in]{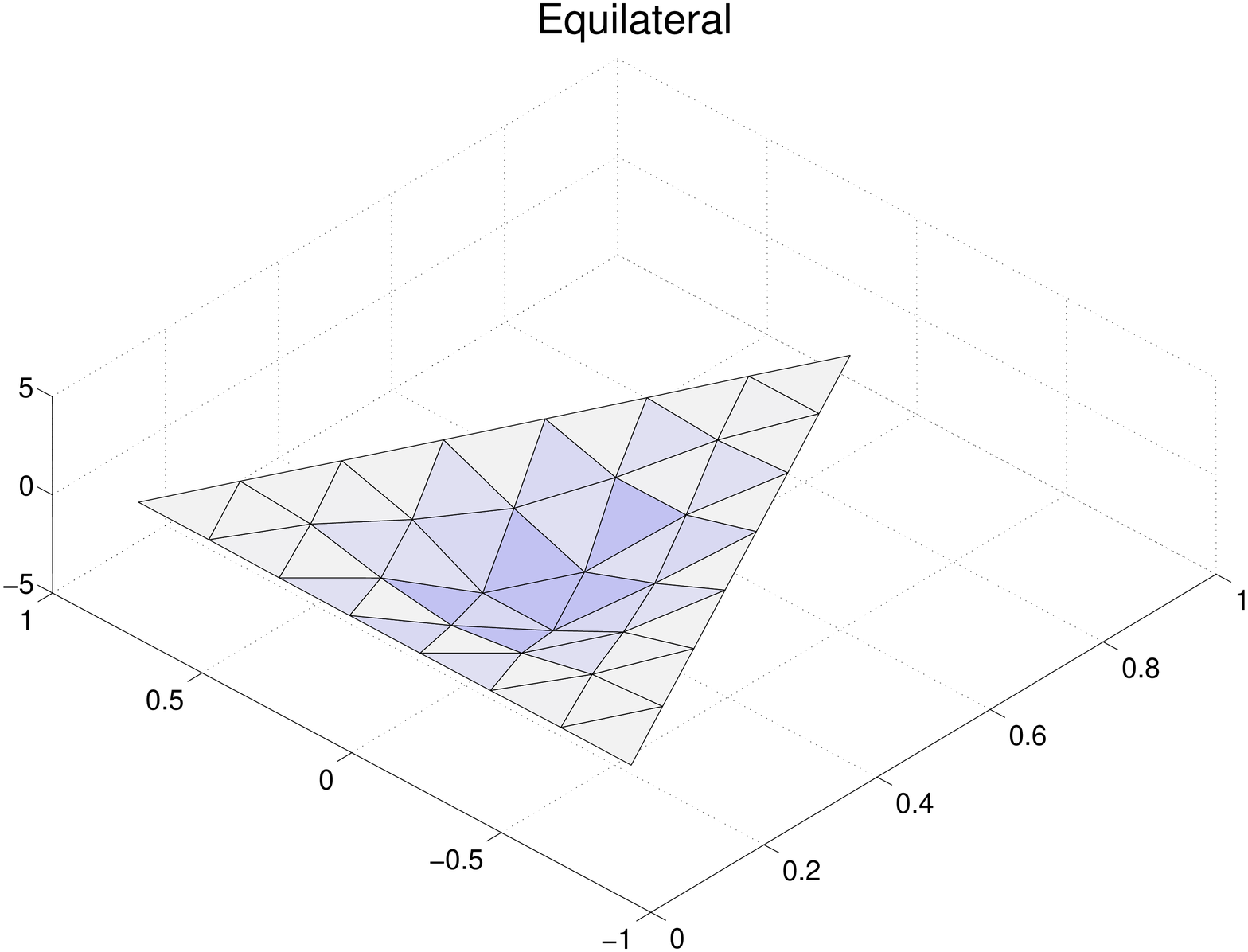} \\
\includegraphics[height=2.3in]{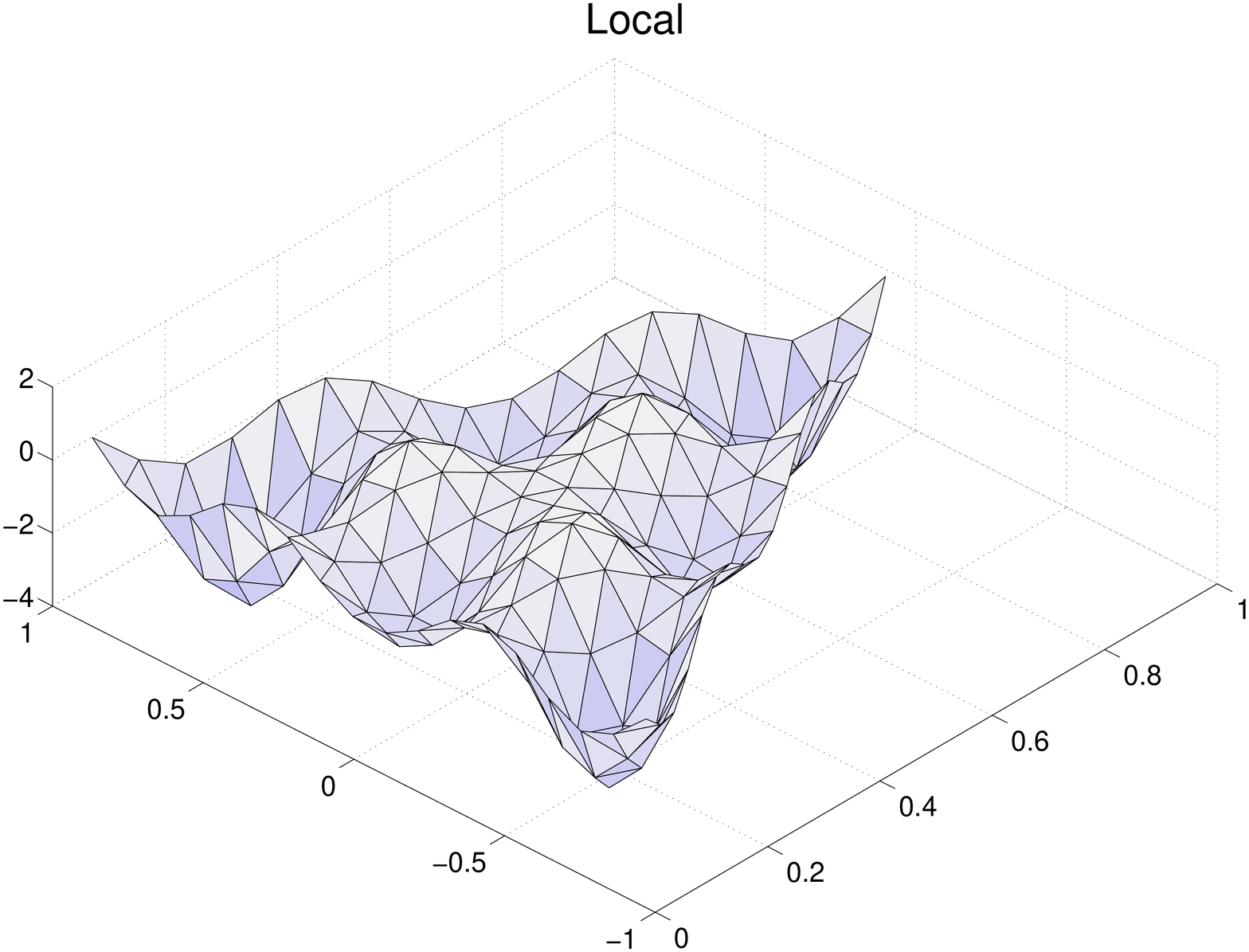} &
\includegraphics[height=2.3in]{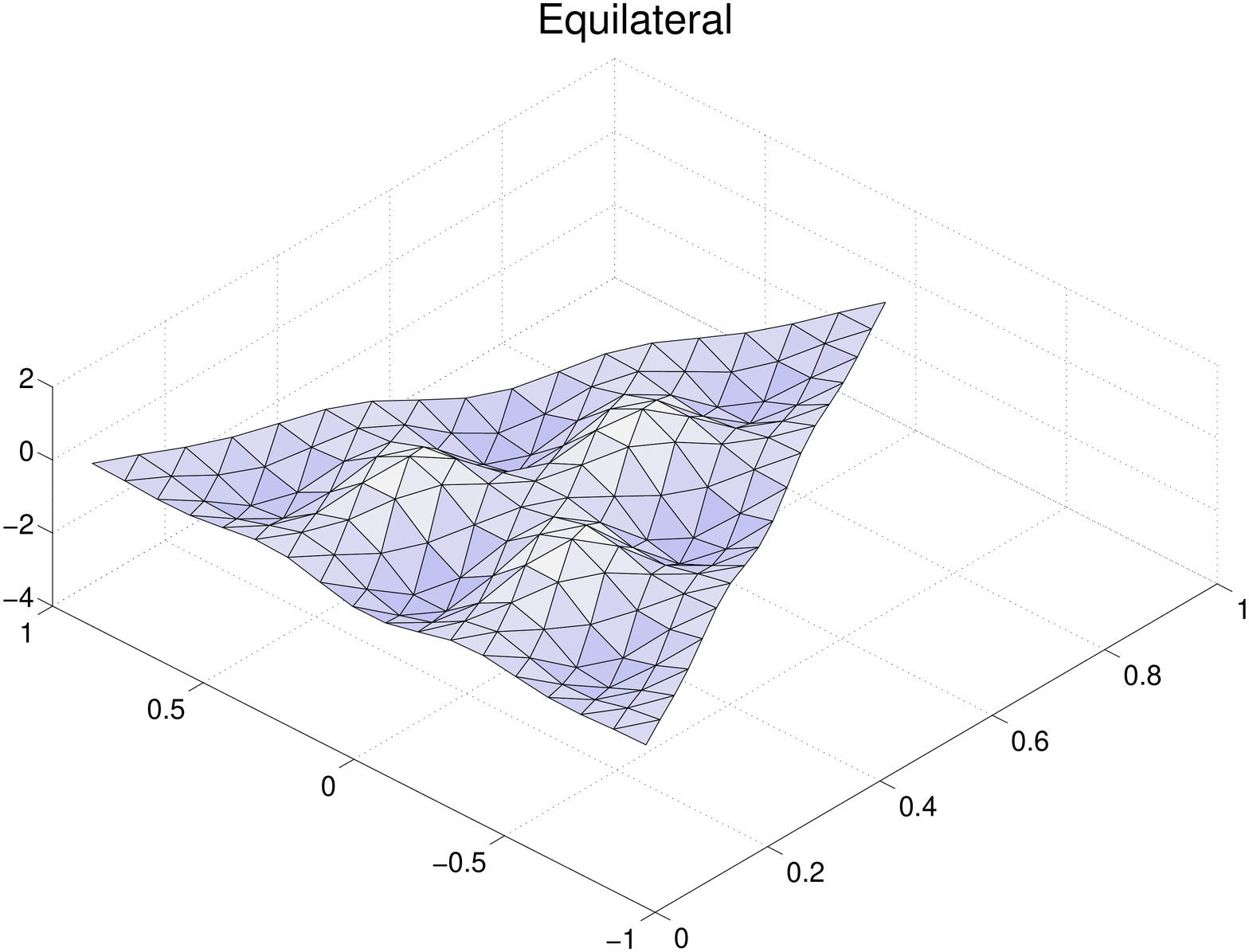} \\
\includegraphics[height=2.3in]{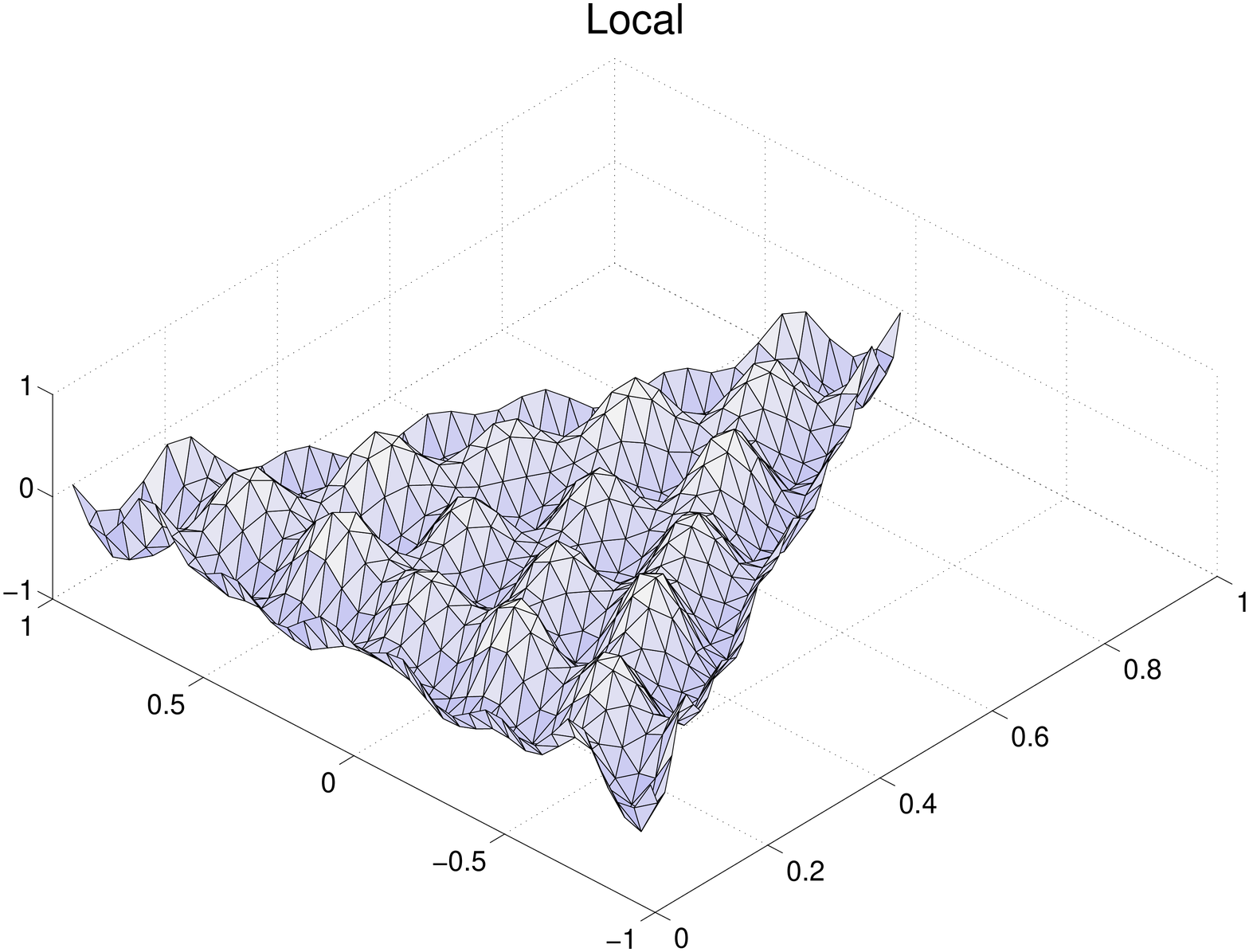} &
\includegraphics[height=2.3in]{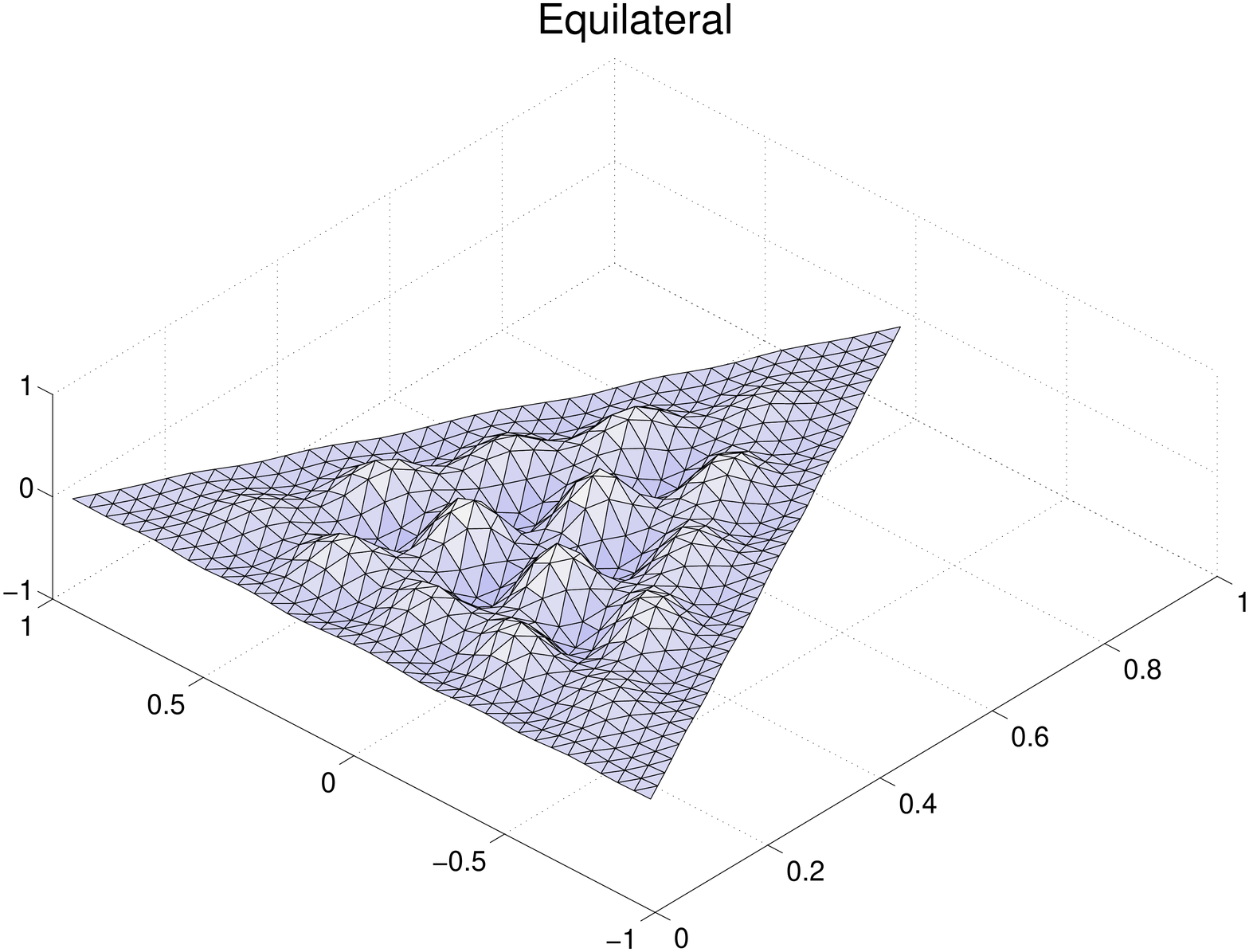} \\
\end{tabular}
\caption{\label{slices}\small Plots of the bispectrum for the local case (on the left) and for the equilateral case (on the right) for $l = \mbox{constant}$ slices using the full radiation transfer function. The top row is $3l = 850$, the second $3l = 1850$, and the third $3l = 3650$.  The bispectrum has been divided by the analytic result so its features are clearly visible.  Lines of constant $l_i$ are those parallel to their respective edge of the triangle.  The extremums of the graphs always appear when the $l_i$ triples contain $l_i$s which represent extremums of the plot along the $l$ direction. Note that the outside of the graphs are suppressed in the equilateral case compared to the local case due to the differing shape functions.}
\end{figure}

For the local case with the full transfer function with an adaptive $N=200$ grid and recursion depth 2 the typical calculation time was 44 mins. In the triangular bispectrum plots  \rfig{slices},
we have $l_1$ constant along the lines parallel to the left bottom edge, where it is at a maximum (400, 900 and 1800 respectively), reducing linearly to a minimum in the top right corner (50 in all three cases).  The dependence on $l_2$ and $l_3$ is similar with respect to the remaining two edges. As we used the analytic version of the local shape function, which has no initial structure, the diagrams are easily understood from figure \ref{bispectrum}.  For $3l=850$, each of the edges corresponds to one $l_i=400$.  As 400 is close to zero in the plot along the equal $l$ direction, the edges are close to zero in the slice.  On the other hand, near  the centre is the triple $(250,\,300,\,300)$  all of which are large and negative so they produce a broad central trough.  In the plot of $3l=1850$ we have a three peaks at the triple $(450,\,700,\,700)$ (plus two permutations) as 700 is large and negative and 450 is positive.  We also have troughs for triples $(250,\,800,\,800)$ (near the corner) and $(500,\,500,\,850)$ (near the middle of the edge) as both 250 and 500 are close to maximums and give strong contributions.  The plot of $3l=3650$ has many peaks and troughs arising from various combinations of the $l_i$'s but they are an order of magnitude smaller than the previous results.  This is because all triples have at least two $l_i \ge 900$ so the contribution from the remaining $l_i$ is suppressed. We can see a 3D comparison in \rfig{3d_local}

\begin{figure}[t]
\centering
\begin{tabular}{cc}
\includegraphics[height=3in]{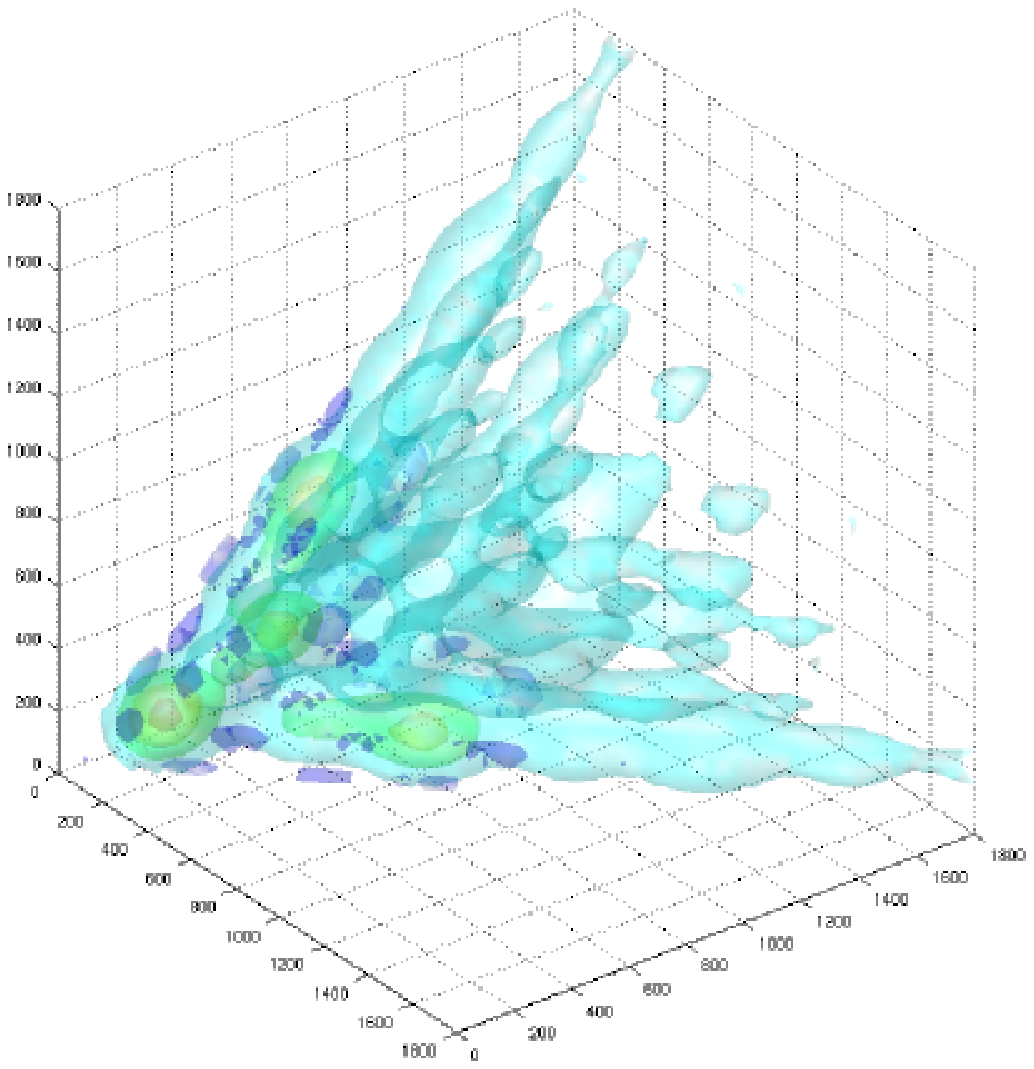} &
\includegraphics[height=3in]{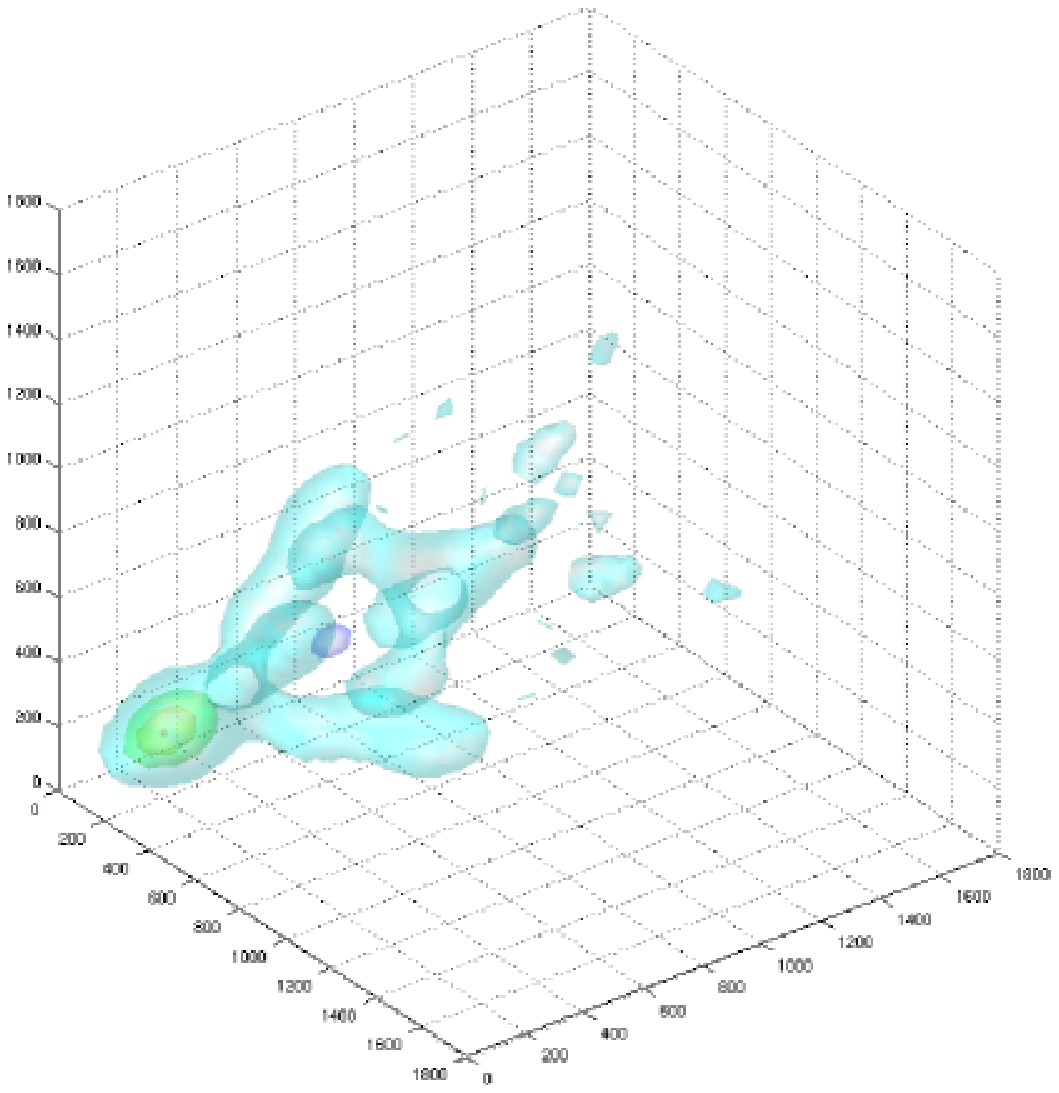}
\end{tabular}
\caption{\label{3d_local}\small Plots of the bispectrum for the local case (on the left) and for the equilateral case (on the right) for $l<1800$.  Note how in the equilateral case all perturbations off the central axis are suppressed}
\end{figure}

For the equilateral case with an initial $N=200$ grid,  the bispectrum calculations were completed on Cosmos again in about 44 mins and are also plotted in figure~\ref{slices} (right).  We see the same basic pattern of peaks and troughs  as in the local case.  The main difference is that the heights of these features become strongly  suppressed towards the edges of the triangle.  This is a direct reflection of the difference in the two shape function shapes. The local shape (superhorizon case) magnifies these features near the edge with strong correlations between very different multipoles, whereas for the equilateral case (horizon-crossing) they are suppressed.

Having proved the method for the two standard analytic cases we can proceed to look at cases where the shape function cannot be separated.  A good example is the DBI model \cite{0404084}.  The shape function for this model is identical to than in \cite{0306122} where higher derivative operators have been added to the Lagrangian.  This shape function has the explicit form of,

\eq
F^{SI}(k_1,k_2,k_3) = \frac{1}{k_1 k_2 k_3 (k_1+k_2+k_3)^2} \(\sum_i k_i^5 + \sum_{i \neq j}(2 k_i^4 k_j - 3 k_i^3 k_j^2) + \sum_{i \neq j \neq l}(k_i^3 k_j k_l - 4 k_i^2 k_j^2 k_l)\).
\qe

\begin{figure}[t]
\centering
\includegraphics[height=2.25in]{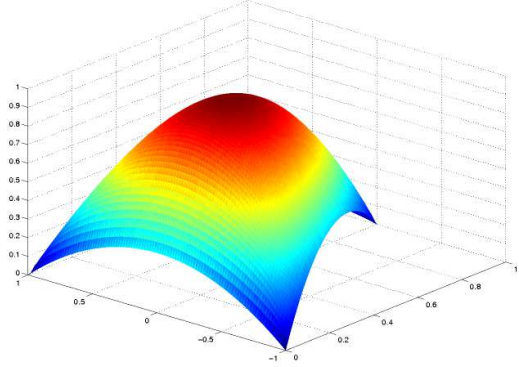}
\caption{\label{primderiv}\small $F^{SI}$ plotted on the $\a\b$-triangle for the `Higher Derivative' model.}
\end{figure}

This shape function has been plotted in figure \ref{primderiv}.

Due to the factor $(k_1+k_2+k_3)^2$ this shape cannot be separated as previously, although some progress has been made by using an integral form \cite{0612571}.  Subsequently we have always been forced to calculate the bispectrum for this model by approximating with the equilateral shape.  With this new approach we can now we can calculate it in full as the non-separability no-longer poses a restriction.  

\begin{figure}[t]
\centering
\includegraphics[height=3.0in]{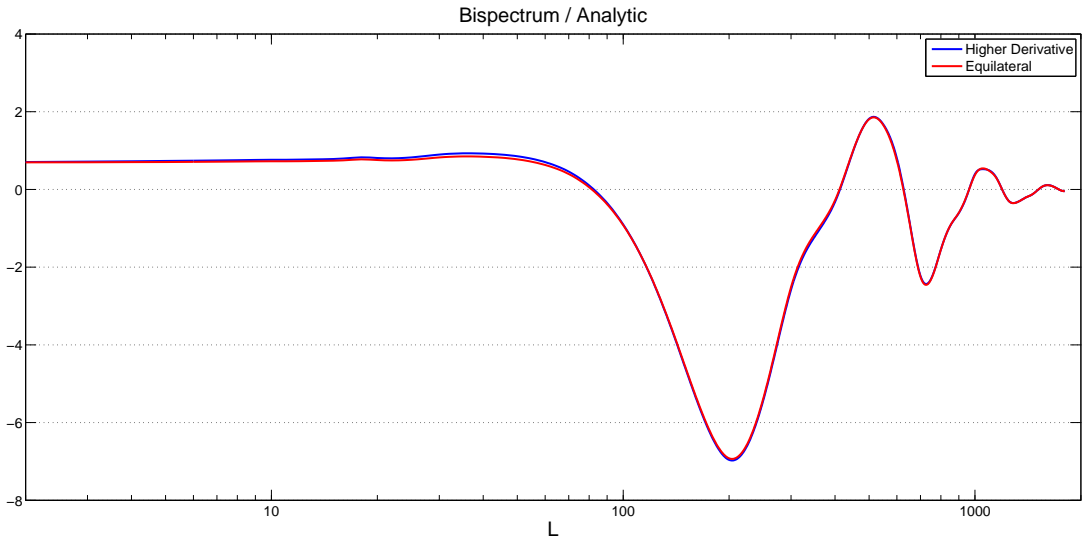}
\caption{\label{bispectrum_de}\small Plot of the bispectrum for the both the higher derivative case and the equilateral case using the full radiation transfer function over the analytic result. As the two shape function shapes are similar in the region surrounding the centre of the triangle the results are almost identical.  To see the difference between the two shapes we must look to the transverse slices.}
\end{figure}

\begin{figure}[t]
\centering
\begin{tabular}{cc}
\includegraphics[height=2.3in]{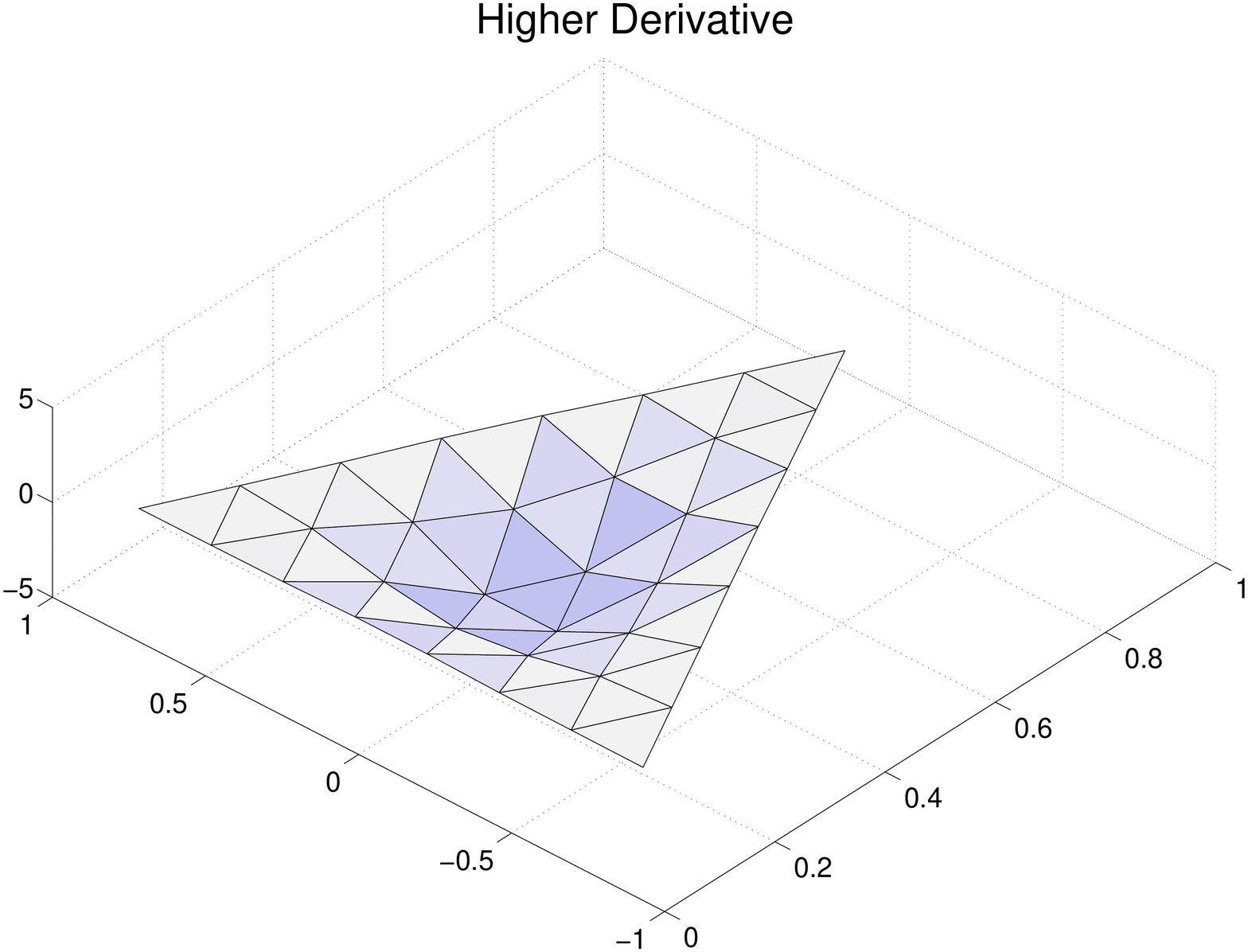} &
\includegraphics[height=2.3in]{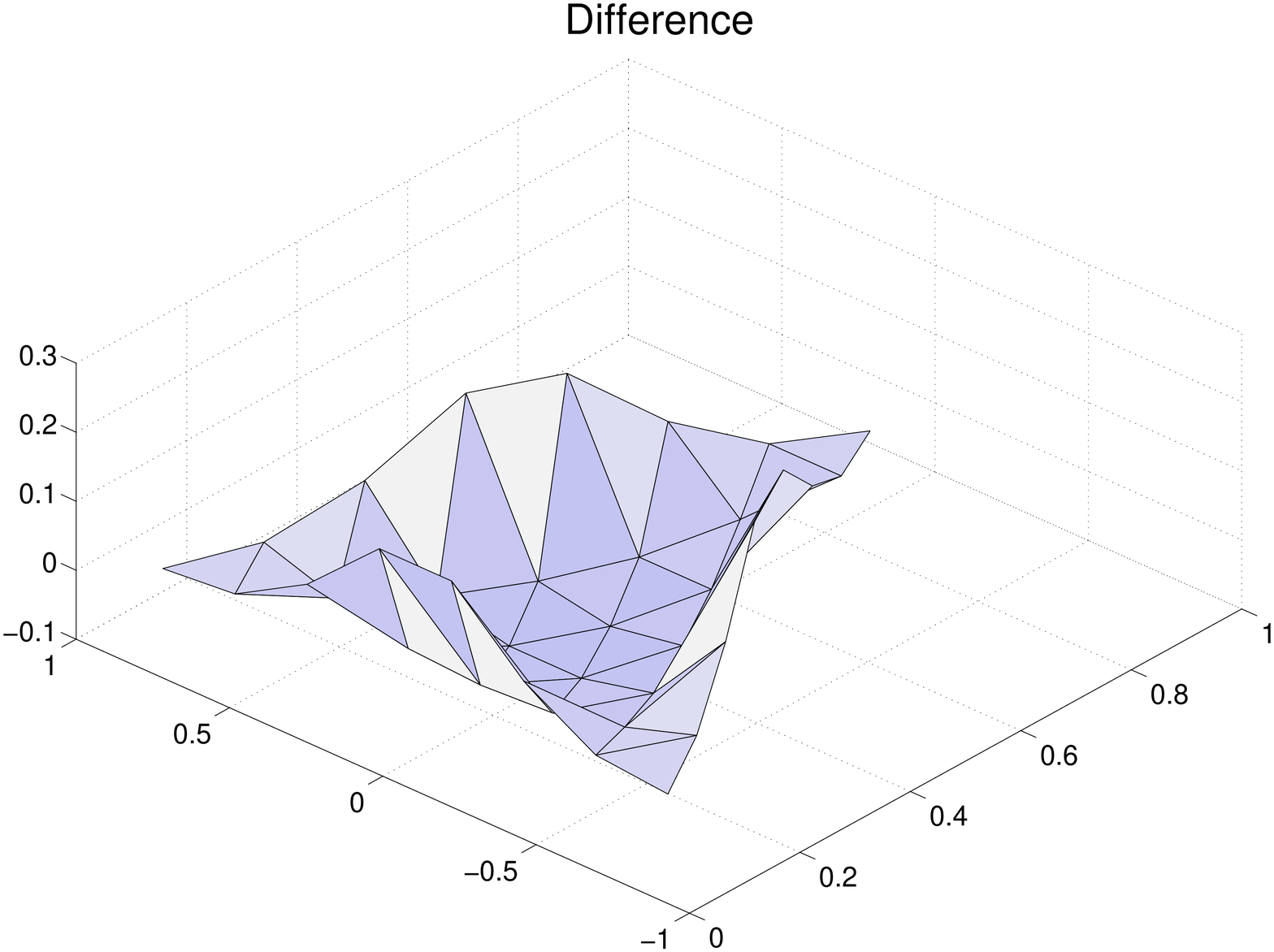} \\
\includegraphics[height=2.3in]{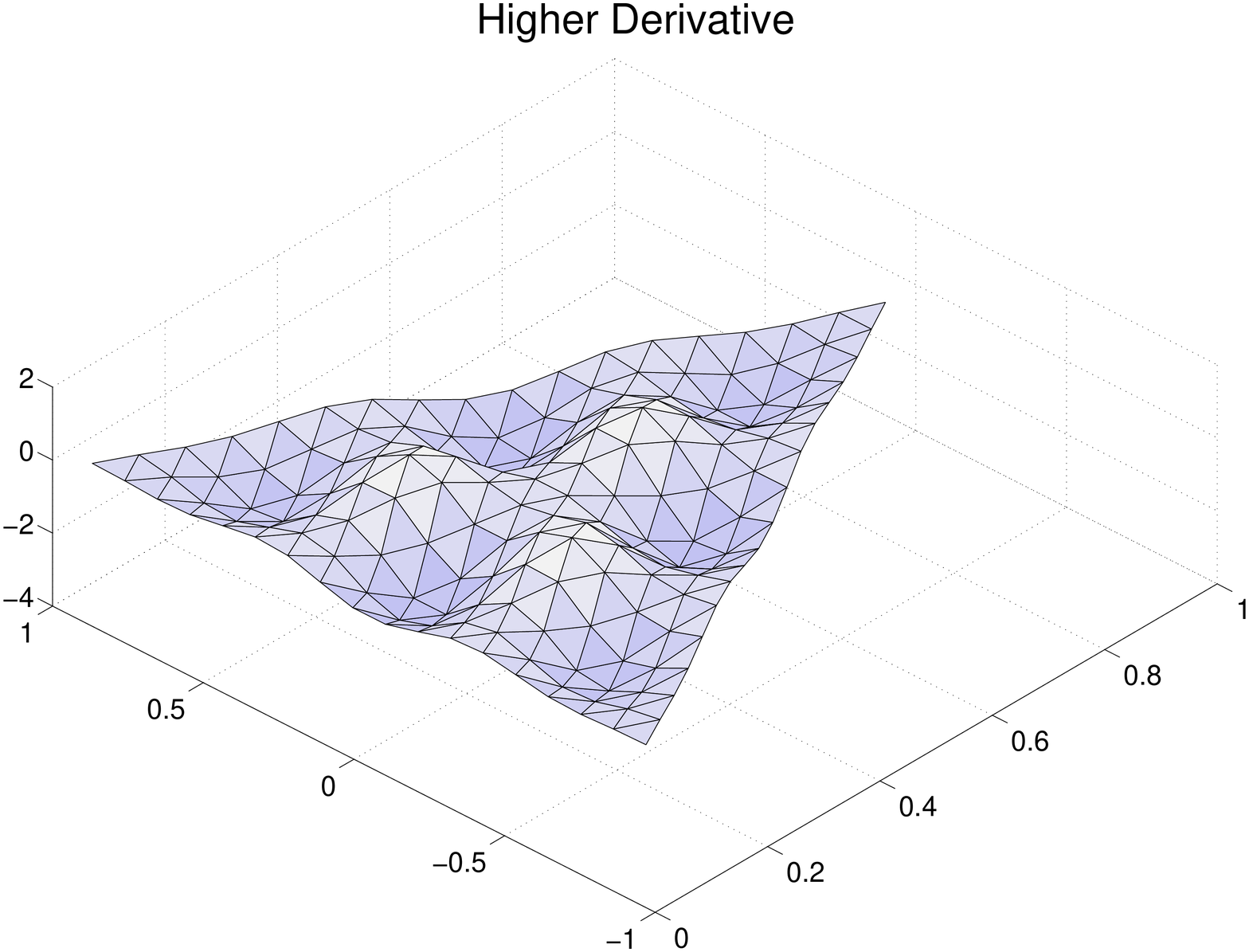} &
\includegraphics[height=2.3in]{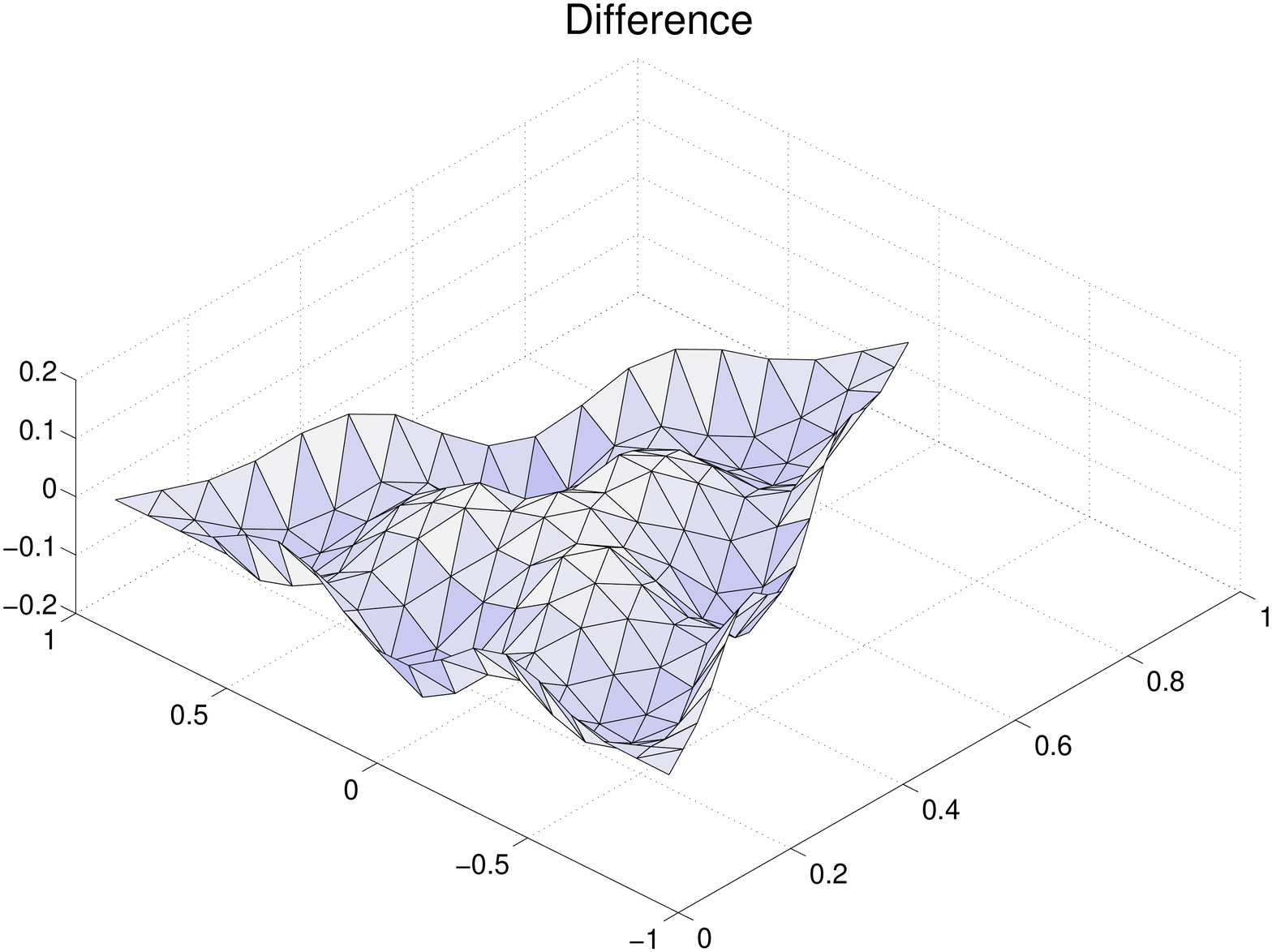} \\
\includegraphics[height=2.3in]{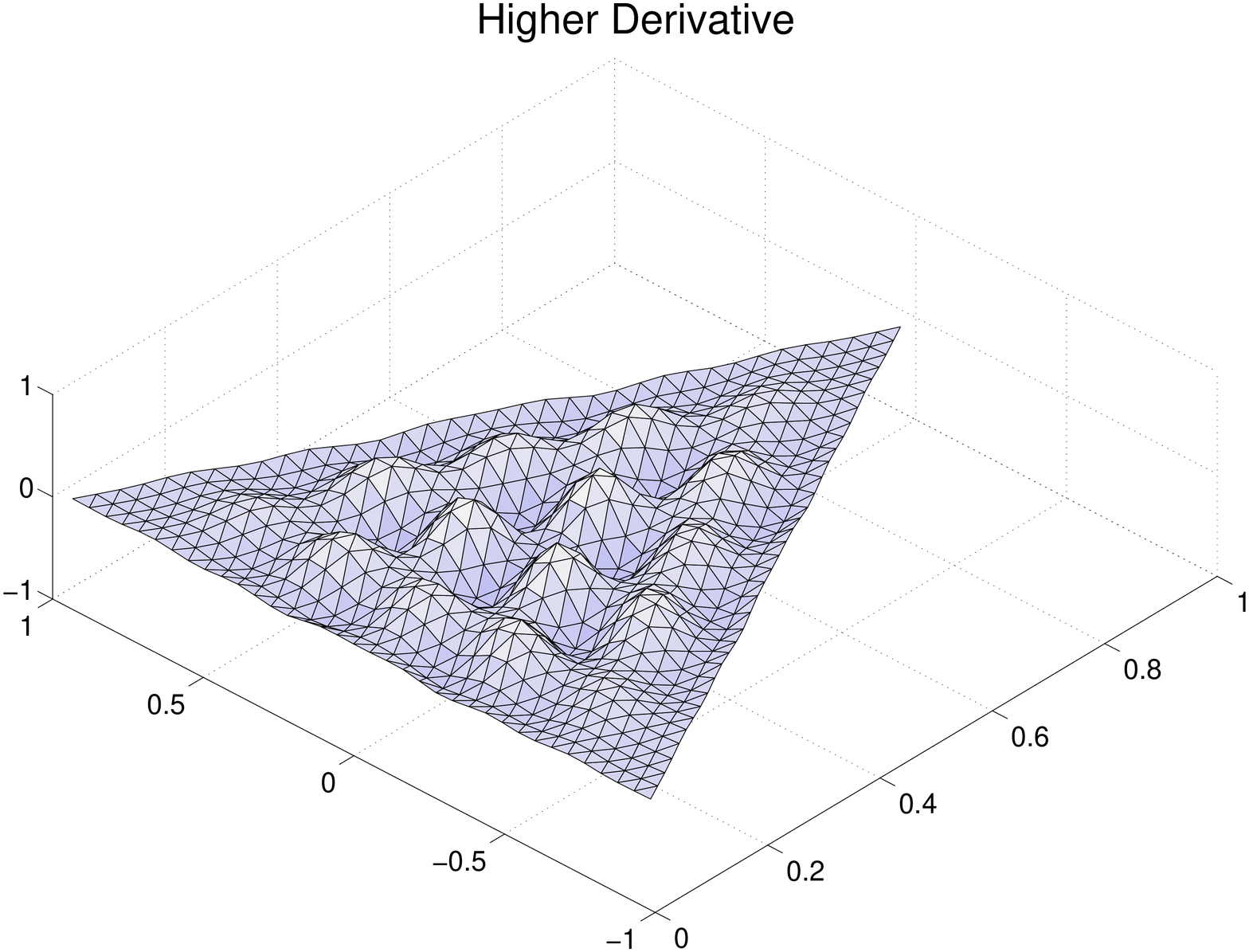} &
\includegraphics[height=2.3in]{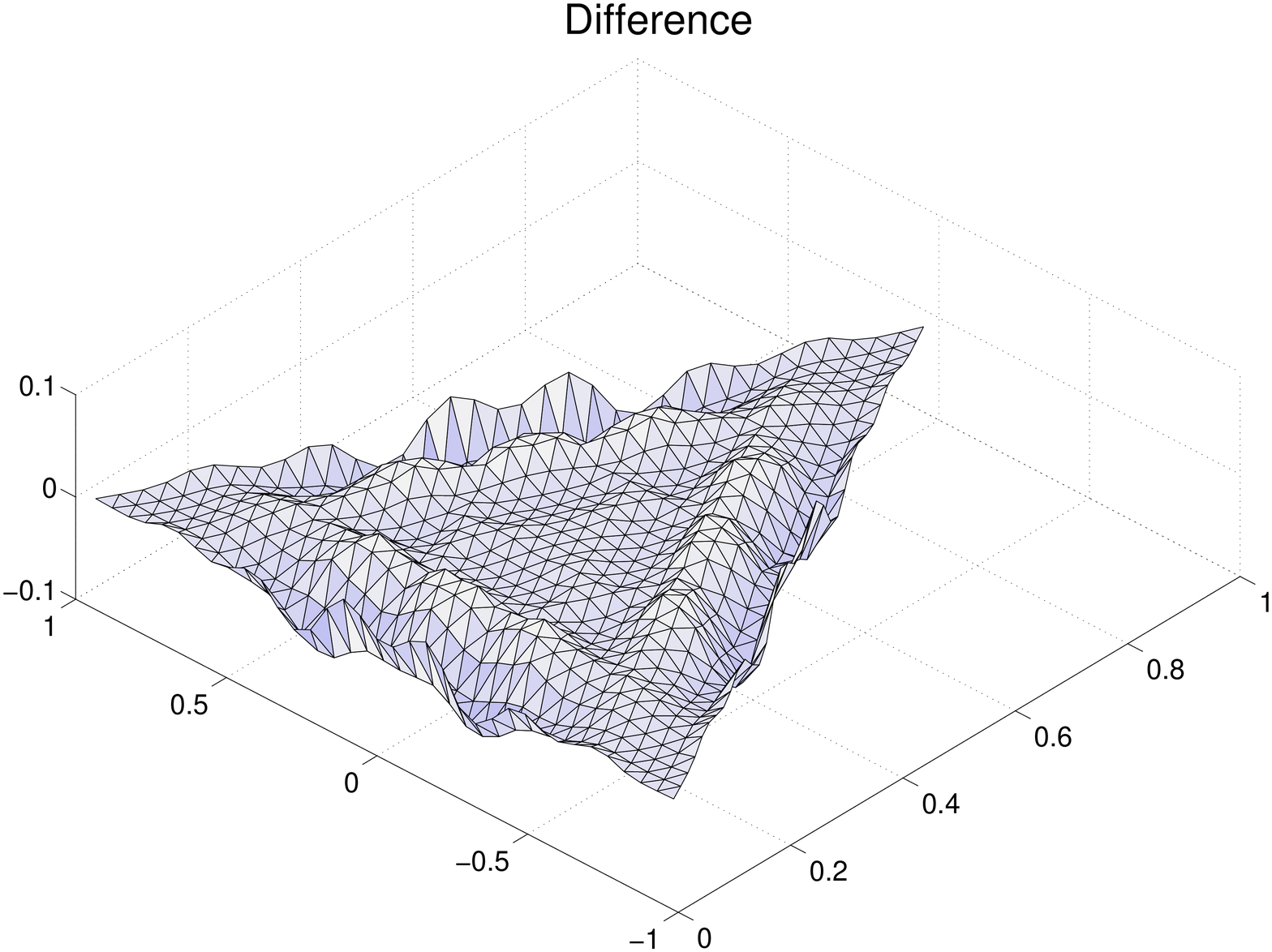} \\
\end{tabular}
\caption{\label{slices2}\small Plots of the bispectrum for the higher derivative model (on the left) and the difference with the equilateral case (on the right) for $l = \mbox{constant}$ slices using the full radiation transfer function. The top row is $3l = 850$, the second $3l = 1850$, and the third $3l = 3650$.  The bispectrum has been divided by the analytic result for the local case so its features are clearly visible.  Lines of constant $l_i$ are those parallel to their respective edge of the triangle.  The extremums of the graphs always appear when the $l_i$ triples contain $l_i$s which represent extremums of the plot along the $l$ direction.}
\end{figure}

The calculations run in similar times as for the equilateral case.  The equal l bispectrum gives almost identical results as the equilateral case as can be seen in \rfig{bispectrum_de} so it may seem that the previous approach is justified. When we plot the transverse slices however, the differences become apparent \rfig{slices2}.  From these plots we can see that the variation between the two cases is at the 10\% level. A 3d comparison can be seen in \rfig{3d_dbi}.

\begin{figure}[t]
\centering
\begin{tabular}{cc}
\includegraphics[height=3in]{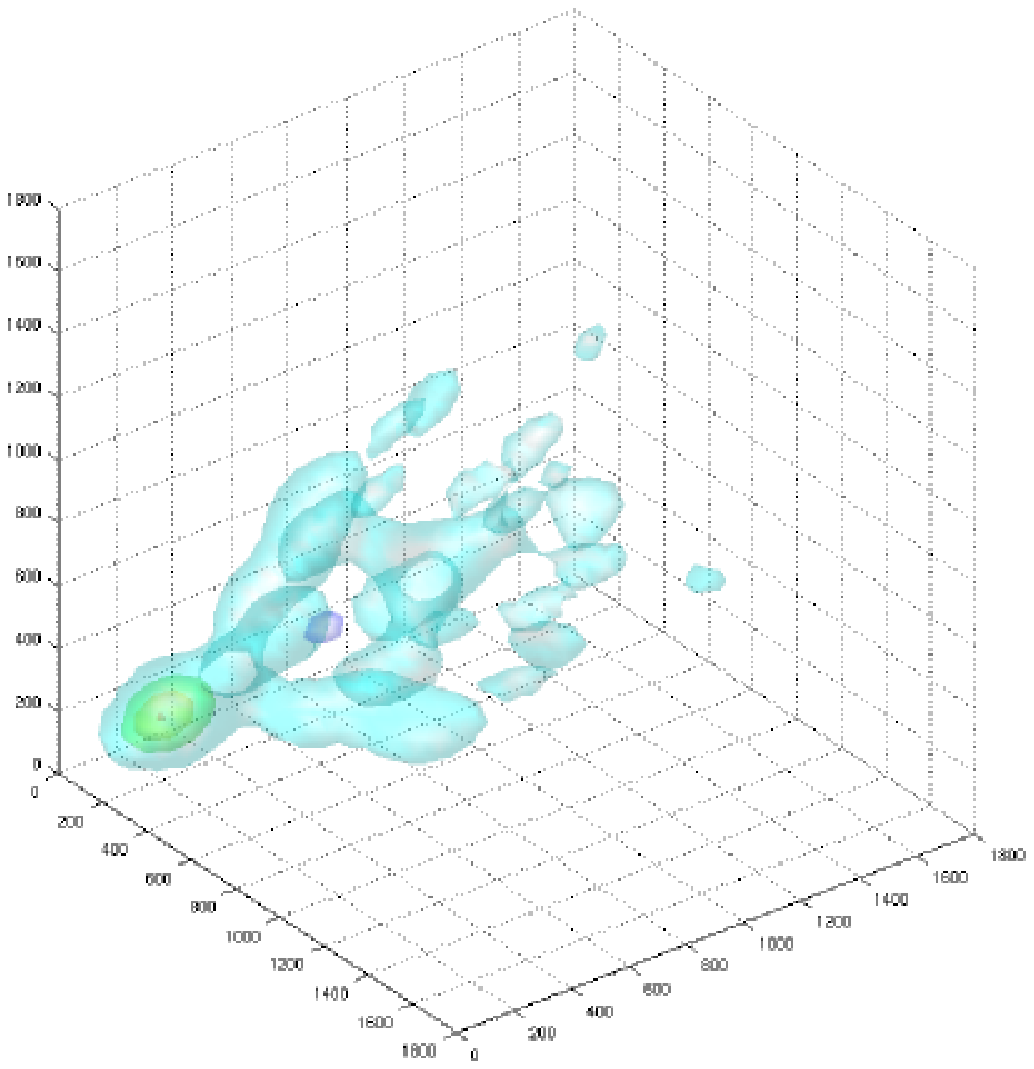} &
\includegraphics[height=3in]{Pictures/Bispectrum_equi.eps}
\end{tabular}
\caption{\label{3d_local}\small Plots of the bispectrum for the DBI case (on the left) and for the equilateral case (on the right) for $l<1800$.  Note how in the equilateral case the perturbations on the sides are suppressed}
\end{figure}

As a final comment, we note that we can exploit the properties of the bispectrum discussed above to speed up 
its evaluation further.   Very substantial improvements could be achieved by tabulating the product of our 1D integrations over the Bessel and transfer functions (\ref{transferint}, \ref{geometricint}), rather than keeping tables of the functions themselves.   This would be relevant, for example, if the background cosmology was fixed (i.e.\ the 
transfer functions are not modified), perhaps for a likelihood analysis marginalising over inflationary model 
parameters. In this case, the table of integrals would have to cover the 5D parameter space ($\a, \,\b,\,l_1,\,l_2,\,l_3$) which may seem unrealistic, but our discussion has shown otherwise.   While the $\a\b$-triangles must be sampled to fairly high resolution (e.g. on an $N=800$ grid), in contrast,  the CMB bispectrum is intrinsically a slowly varying function of the multipoles (period $l\sim 200$, like the CMB power spectrum).  Provided the shape function is relatively featureless, we would only require sparse sampling in multipole space (perhaps as few as 33 points on a 2D slice, given the underlying symmetries, even for $l \sim 1000$).
The initial tabulation would entail much parallel computation,  but 
subsequently complete bispectra (up to $l < 2000$) could be accurately calculated in minutes.

\section{Estimation}

Having calculated the bispectrum accurately today we then need to turn to the problem of measurement.  The bispectrum itself will unfortunately be too small for direct measurement so we will have to rely on estimators.  Optimal estimators for the bispectrum are based on the statistic,

\eq
S = \frac{1}{N} \sum_{l_i} \frac{1}{C_{l_1} C_{l_2} C_{l_3}} B_{l_1 l_2 l_3} \sum_{m_i} \( \begin{array}{ccc} l_1 & l_2 & l_3 \\ m_1 & m_2 & m_3 \end{array} \) a_{l_1 m_1} a_{l_2 m_2} a_{l_3 m_3},
\qe

where $B_{l_1 l_2 l_3}$ is the bispectrum predicted from theory and the $a_{l m}$'s are the measured values.  If the normalisation factor is,

\eq
N = \sum_{l_i} \frac{B^2_{l_1 l_2 l_3}}{C_{l_1} C_{l_2} C_{l_3}},
\qe

\noindent then the statistic gives the relative magnitude of the theoretical bispectrum that best fits the data.  With the new methods described in the previous section we can now calculate the theoretical bispectrum accurately for an almost arbitrary shape function. Unfortunately however, we are still be restricted by difficulties in calculating the observed bispectrum.  One of the biggest challenges is to calculate the Wigner 3j-symbols.  These have no simple analytic form for the general case and are too many too tabulate for large values of l.  Even for a relatively modest $l_{max} = 335$ as used for analysis of the WMAP 1-year data requires calculating over 80 billion independent 3j-symbols. This makes the sum over $m$ impractical to compute even if the $a_{lm}$'s were all known.  The separable case overcomes this by using the Gaunt integral to replacing the 3j-symbol with an integral over three spherical harmonics which are absorbed with the $a_{lm}$'s into the separable parts allowing for efficient calculation.  In the general case this is impossible so we must find another way around the problem.

\begin{figure}[t]
\centering
\includegraphics[height=3.0in]{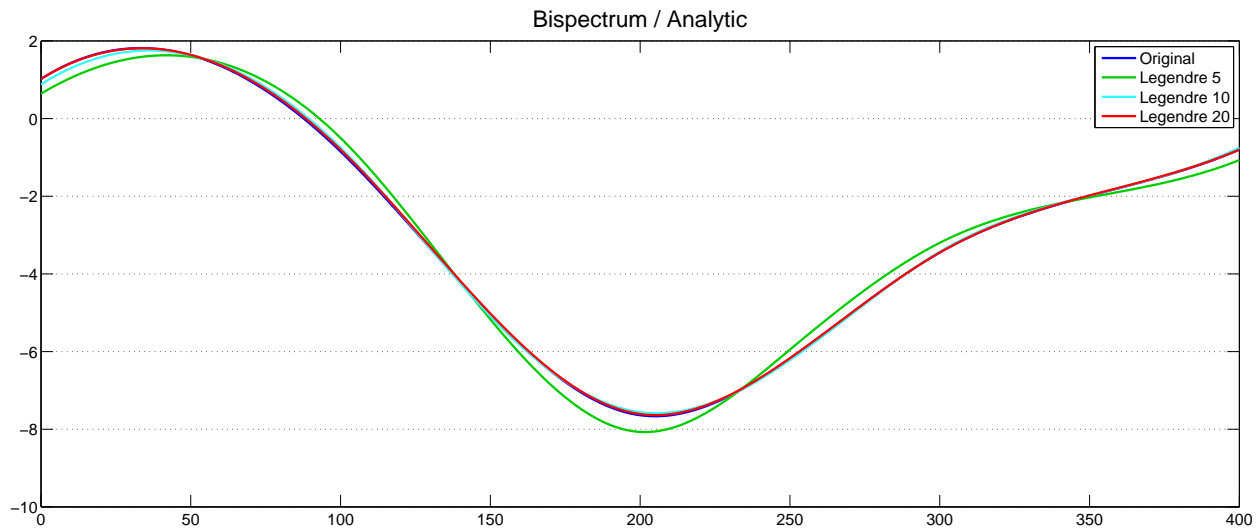}
\caption{\label{recon}\small Plot comparing the equal $l$ reduced bispectrum and the representation of it in terms of basis functions up to $\a = 30$}
\end{figure}

\begin{figure}[t]
\centering
\begin{tabular}{cc}
\includegraphics[height=2.3in]{Pictures/bispectrum850.eps} &
\includegraphics[height=2.3in]{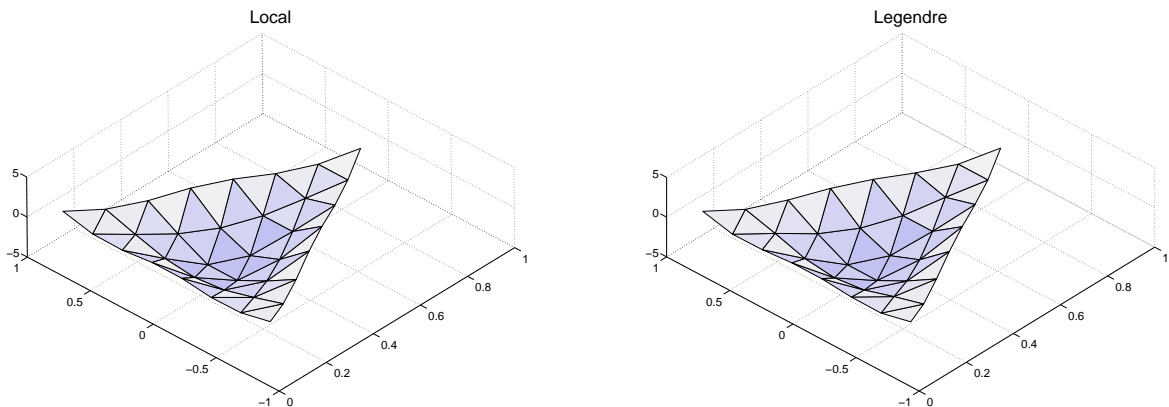}
\end{tabular}
\caption{\label{reconslice}\small Plots of the bispectrum for the local model plotted next to the representation of it in terms of basis functions up to $\a = 30$.}
\end{figure}

In \cite{0612571} a fast method is proposed for calculating the estimator. When the reduced bispectrum can be represented in a separable form,
\eq
b_{l_1 l_2 l_3} = \frac{1}{6} \sum^{N_{fact}}_{i=1} \(X^{(i)}_{l_1} Y^{(i)}_{l_2} Z^{(i)}_{l_3} + \mbox{5 permutations}\).
\qe
Then if we defining,
\eq
X^{(i)}_a(\un) = \sum_{lm} X^{(i)}_l \frac{a_{lm}}{C_{l}} Y_{lm}(\un),
\qe
we can write the estimator as,
\eq
S = \frac{1}{N} \sum^{N_{fact}}_{i=1} \int d\un X^{(i)}_a(\un) Y^{(i)}_a(\un) Z^{(i)}_a(\un).
\qe
While the general reduced bispectrum has no simple separable form, we note from our earlier plot that it is smooth. This means that we should be able to represent it as a sum of smooth basis functions. As we have seen from previous plots (\ref{bispectrum_eq}, \ref{slices}, \ref{slices2}) when we divide the bispectrum by the analytic result we get a surface which is $\mbox{O}(1)$ with slow oscillations. As a result we choose our basis as follows,
\eq
b_{l_1 l_2 l_3} = \frac{1}{3} \sum_{\a \b \g} a_{\a \b \g} \(X'_{\a}(l_1) X'_{\b}(l_2) X_{\g}(l_3)+ \mbox{2 permutations}\),
\qe
where $X_{\a}$ are shifted Legendre polynomials,
\eq
X_{\a}(l) = P_{\a}(\frac{2l-l_{max}}{l_{max}})
\qe
and,
\eq
X'_{\a}(l) = \frac{X_{\a}(l)}{l(l+1)}.
\qe
However, any set of orthogonal polynomials would suffice. Simplifying we have
\eq
\(\frac{l_1(l_1+1)l_2(l_2+1)l_3(l_3+1)}{l_1(l_1+1) + l_2(l_2+1) + l_3(l_3+1)}\)b_{l_1 l_2 l_3} = \sum_{\a \b \g} a_{\a \b \g} X_{\a}(l_1) X_{\b}(l_2) X_{\g}(l_3)
\qe
And $a_{\a \b \g}$ can be easily determined,
\eq
a_{\a \b \g} = (2\a+1)(2\b+1)(2\g+1)\int\frac{d l_1 d l_2 d l_3}{l_{max}^3}\(\frac{l_1(l_1+1)l_2(l_2+1)l_3(l_3+1)}{l_1(l_1+1) + l_2(l_2+1) + l_3(l_3+1)}\)b_{l_1 l_2 l_3}X_{\a}(l_1) X_{\b}(l_2) X_{\g}(l_3),
\qe
using the orthogonality condition,
\eq
\int^1_{-1} dx P_{\a}(x) P_{\b}(x) = \frac{2\d_{\a\b}}{2\a + 1} \imp \int^{l_{max}}_0 \frac{dl}{l_{max}} X_{\a}(l) X_{\b}(l) = \frac{\d_{\a\b}}{2\a + 1}.
\qe
If we define as before,
\eq
\bar{X}_{\a}(\un) = \sum_{lm} X_{\a}(l) \frac{a_{lm}}{C_{l}} Y_{lm}(\un),
\qe
and
\eq
\bar{X}'_{\a}(\un) = \sum_{lm} \frac{X_{\a}(l)}{l(l+1)} \frac{a_{lm}}{C_{l}} Y_{lm}(\un),
\qe
Then the estimator becomes,
\eq
S = \frac{1}{N} \sum_{\a \b \g} a_{\a \b \g} M_{\a \b \g}.
\qe
Where we have defined
\eq
M_{\a \b \g} =  \frac{1}{3} \int d\un \(\bar{X}'_{\a}(\un) \bar{X}'_{\b}(\un) \bar{X}_{\g}(\un) + \mbox{2 permutations} \).
\qe
This approach has several advantages.  Firstly, there is complete separation of theory and measurement.  The quantities $\bar{X}$,$\bar{X'}$ and $M_{\a \b \g}$ only need to be calculated once per map then $M_{\a \b \g}$ can simply be stored on disk.  For each theory we only need to calculate $a_{\a\b\g}$ then perform the sum of the product of it with the waiting $M_{\a \b \g}$.  This process will be quick as rather than summing over all combinations of $l_i$ we now only need to do so over the range of $\a,\b,\g$.  This new approach for the estimator makes it possible to produce quick and accurate bispectrum estimates for general models.
This decomposition has been tested for $l_{max} = 400$ for the local model.  Using basis functions up to $\a = 30$ we managed to reconstruct the reduced bispectrum to a 1\% accuracy as shown by the plots (\ref{recon}, \ref{reconslice}).

\section{Future work}

With the numerical methods we have developed it is now possible to accurately calculate the CMB bispectrum today from a general shape function.  In the stringent test case there is a problem obtaining accuracy without long calculation times because of truncation errors, but arbitrary accuracy can be achieved by altering truncation thresholds and improving resolution.  In the real calculation using the full radiation transfer function we have demonstrated that the systematic error sources are much weaker and very accurate results can be obtained rapidly using our adaptive method.  We find that the bispectra for the local and equilateral cases are broadly similar in the equal multipole case, with the main difference being that the equilateral case is smaller at low $l$.  Where the two cases differ significantly is in the slices  $l_1 + l_2 + l_3 = \mbox{contant}$ (and for which we find no significant new sources of error).  For the calculation with the full radiation transfer functions, we find that the results are much the same in the centre, but with the 
equilateral case more heavily suppressed where the $l$ values differ significantly, as expected. This indicates that the equal multipole bispectrum provides a good measure of the overall size of the non-Gaussianity, while the unequal $l$ bispectrum provides a differentiator between competing models. Obviously, meaningful plots of the CMB bispectrum (like figure \ref{slices}) are not expected to be determined observationally, not least because they will be dominated by noise and convolved with experimental beams. Instead we must use an estimator to determine the fit of theory to observation. 

While a parallel implementation of the code runs sufficiently rapidly on a supercomputer for calculations of the full bispectrum, there is still plenty of scope for further improvements.  Essentially all of the calculation time is spent evaluating the one-dimensional integrals with the geometric integral taking twice as long as the transfer integral.  To reduce overall evaluation times we then have two options.  We can improve the integration methods for the one-dimensional integrals or we can reduce the number of points we need to calculate by improving the adaptive algorithm.  Improving the cell selection criteria in the adaptive algorithm has largely been a long process of trial and error, and no doubt this  can be continued with further success.   However,  improving the one-dimensional integration method is perhaps the more exciting option.  Here, more rapidly convergent numerical methods are possible, but the hope is that for the case of the geometric integral we may be able to switch to an analytic solution should a stable method of evaluation be discovered.  Such an advance could reduce overall bispectrum calculation times by a factor of 3.

The decomposition of the bispectrum into Legendre polynomials has also been parallelised but is still hampered but the requirement to use a large number of polynomials.  This arises from having to perform the integral over the cube in l space with sides from $0$ to $l_{max}$ while the bispectrum is only defined inside the tetrahedron created by the triangle condition on the $l$'s.  When we use the code to calculate points outside this tetrahedron we introduce discontinuities which require a large number of polynomials to fit accurately. We are currently looking to edit the code to calculate these points so that the bispectrum remains smooth as we cross the boundary.

The present code has been developed in modular form to link and work together with existing CMB line-of-sight codes for the angular power spectrum (like CMBFAST, CAMB or CMB2000); it extracts the appropriate radiation transfer functions from these for a given cosmology.   In due course, it is our intention to improve the efficiency, portability and parallelism (as well as the documentation) of this code and to make it publicly available.

The bispectrum remains a highly significant tool for testing for non-Gaussianity in the CMB. Hopefully, future experiments like the Planck satellite will either detect or provide much tighter constraints on this non-Gaussianity, opening a new window on the early universe.   With this code we now have the capability to 
make accurate predictions for the contributions to the CMB bispectrum which are of primordial origin.
For example, we plan to use quantitative bispectra from multifield inflation models \cite{0511041} and project these forward to make falsifiable CMB predictions.   
Clearly parallel efforts must also continue in order to pin down and distinguish alternative sources of CMB non-Gaussianity from later times, such as second-order effects in the radiation transfer functions and nonlinear astrophysical effects.  

\section{Acknowledgements}
We are especially grateful to Kendrick Smith for conversations regarding the decomposition of the reduced bispectrum into polynomials and the subsequent method for calculating the estimator. We are also grateful for useful discussions with Paolo Creminelli, Michele Liguori, Gerasimos Rigopoulos and Bartjan van Tent.   Simulations were performed on the COSMOS supercomputer (an Altix 3700) which is funded by PPARC, HEFCE and SGI.   We are very grateful for help from the COSMOS programmer, Victor Travieso, who parallelised the code and improved its efficiency.  The CMB2000 code, from which the radiation transfer functions were extracted, was written by Martin Landriau, EPS, Martin Bucher and Richard Battye.   This research was supported by PPARC grant PP/C501676/1, the Sims Fund, Churchill College, the Cambridge Commonwealth trust, a Edward and Isabel Kidson Scholarship and a William Georgetti Scholarship.

\appendix

\section{Three Bessel function integrals}

Higher-order angular correlation functions generically involve integrals over products of Bessel 
functions.  
If we allow a general bispectrum but remain in the large angle approximation then both the 1D integrals in (\ref{biint2}), $I^T(\a,\b)$ and $I^G(\a,\b)$, are geometric and of this form. Assuming $n=0$,
\eq
I^T(\a,\b) \= \frac{1}{3^3} \int j_{l_1}\(a k\) j_{l_2}\(b k\) j_{l_3}\(c k\) \frac{dk}{k},
\qe                                                                                                         
where $\D\n$ has been absorbed via a rescaling of $k$.  This leaves us trying to solve integrals of the form,
\eq
I_{l_1,l_2,l_3}(a,b,c,n) = \int^\infty_0 j_{l_1}\(a x\) j_{l_2}\(b x\) j_{l_3}\(c x\) x^n dx
\qe
where $n=2,\,-1$.

We already have a several solutions for $n=2$ given in refs.~\cite{9107011,9309023,0506114}.  For the general case a solution is proposed in ref.~\cite{8907204} but,  unfortunately, their analysis is incomplete.  Their method revolves around replacing the spherical Bessel functions as below
\eq
j_{l}(x) = \sum_{m=0}^l \frac{(l+m)!}{(l-m)!m!}\(\frac{1}{2x}\)^{m+1}\(e^{-ix}(i)^{l-m+1} + e^{ix}(-i)^{l-m+1}\)
\qe
to rewrite the integral as,
\eq \label{threebesselint}\nonumber &&\sum_{m_1=0}^{l_1} \sum_{m_2=0}^{l_2} \sum_{m_3=0}^{l_3} \(\prod^3_{i=1} \frac{(l_i+m_i)!}{(l_i-m_i)!m_i!} \) \frac{1}{(2a)^{m_1+1}(2b)^{m_2+1}(2c)^{m_3+1}} \int_0^{\infty} \frac{(i)^{l_1+l_2+l_3-m_1-m_2-m_3+3}}{x^{m_1+m_2+m_3-n+3}}\\
&&\(e^{-iax} + e^{iax}(-1)^{l_1-m_1+1}\) \(e^{-ibx} + e^{ibx}(-1)^{l_2-m_2+1}\) \(e^{-icx} + e^{icx}(-1)^{l_3-m_3+1}\) dx.  \qe
The second line of the integrand can be expanded to,
\eq
(i)^{l_1+l_2+l_3-m_1-m_2-m_3+3} &&\[ e^{-i(a+b+c)x} + e^{i(a+b+c)x}(-1)^{l_1+l_2+l_3-m_1-m_2-m_3+3} \right. \\
\nonumber &&+ (-1)^{l_1-m_1+1} \( e^{-i(-a+b+c)x} + e^{i(-a+b+c)x}(-1)^{l_1+l_2+l_3-m_1-m_2-m_3+3} \) \\
\nonumber &&+ (-1)^{l_2-m_2+1} \( e^{-i(a-b+c)x} + e^{i(a-b+c)x}(-1)^{l_1+l_2+l_3-m_1-m_2-m_3+3} \) \\
\nonumber &&+ \left.(-1)^{l_3-m_3+1} \( e^{-i(a+b-c)x} + e^{i(a+b-c)x}(-1)^{l_1+l_2+l_3-m_1-m_2-m_3+3} \) \].
\qe
So we are trying to complete the integral \ref{threebesselint} for eight values of $\l$,
\eq
\int_0^{\infty} \frac{e^{i \l x}}{x^P}dx.
\qe
This integral is singular but the sum is not.  The authors calculate,
\eq
\int_0^{\infty} \frac{e^{i \l x}}{(x+\e)^P}dx,
\qe
and assert that the infinite terms cancel explicitly in the sum as $\e \to 0$, which is incorrect.  If we instead calculate,
\eq
\int_{\e}^{\infty} \frac{e^{i \l x}}{x^P}dx.
\qe 
which has solution,
\eq \frac{(i \l )^{P-1} }{ (P-1)! } E_1(-i \e \l) + e^{i \e \l}\sum_{s=0}^{p-2}\frac{(p-2-s)!}{(p-1)!} \frac{(i\l)^s}{\e^{p-1-s}}, \qe
where $E_1$ is the En-Function with series expansion,
\eq E_1(-i \e \l) = - \g - \ln(\e) - \ln | \l | + \frac{ i \pi }{ 2 } \mbox{sgn}( \l ) + O(\e), \qe
and $\g$ is the Euler constant.  This is identical to that in ref.~\cite{8907204} result except for the $e^{i \e \l}$ factor in front of the sum.  The  claim that the infinite part cancelled led the authors to drop the sum altogether. Here the singular parts of the sum over $s$ cancel in the triple sum over $m_i$, and as we will set $\e$ to zero, we are then only  left with in the constant part of the series.
\eq e^{i \e \l}\sum_{s=0}^{p-2}\frac{(p-2-s)!}{(p-1)!} \frac{(i\l)^s}{\e^{p-1-s}} = \sum^{\infty}_{n=0}\frac{(i \e \l)^n}{n!}  \sum_{s=0}^{p-2}\frac{(p-2-s)!}{(p-1)!} \frac{(i\l)^s}{\e^{p-1-s}}. \qe
Which gives,
\eq \sum_{s=0}^{p-2} \frac{(i \e \l)^{p-1-s}}{(p-1-s)!} \frac{(p-2-s)!}{(p-1)!} \frac{(i\l)^s}{\e^{p-1-s}} &=& \sum_{s=0}^{p-2} \frac{(i \l)^{p-1}}{p-1-s} \frac{1}{(p-1)!}\\
\nonumber &=&  \frac{(i \l)^{p-1}}{(p-1)!} \sum_{s=0}^{p-2} \frac{1}{p-1-s} \\
\nonumber &=&  \frac{(i \l)^{p-1}}{(p-1)!} \mbox{H}_{p-1}.
\qe
Where $\mbox{H}_a=\sum_{s=1}^{a} \frac{1}{s}$ is the harmonic number.

In the triple sum over $m_i$ the parts proportional to $\g$ and $\ln(\e)$ in $E_1$ cancel exactly. The part of the integral that contributes to the final result is then,
\eq \frac{(i \l )^{P-1}}{(P-1)!} \( \mbox{H}_{p-1} - \ln | \l | + \frac{ i \pi }{ 2 } \mbox{sign}( \l ) \). \qe
We can then substitute this into the expression for the integral over the spherical Bessel functions to eventually obtain,
 
\eq \nonumber && (i)^{l_1+l_2+l_3-n}\sum_{m_1=0}^{l_1} \sum_{m_2=0}^{l_2} \sum_{m_3=0}^{l_3}\frac{1}{(m_1+m_2+m_3-n+2)!}\(\prod^3_{i=1}(-1)^{m_i}\frac{(l_i+m_i)!}{(l_i-m_i)!m_i!}\) \Bigg\{ \\
\nonumber && i (1+(-1)^{l_1+l_2+l_3+n+1}) \[ F_1(a,b,c) + (-1)^{l_1} F_1(-a,b,c) + (-1)^{l_2} F_1(a,-b,c) + (-1)^{l_3} F_1(a,b,-c)  \]\\
\nonumber && + \frac{\pi}{2}(1+(-1)^{l_1+l_2+l_3-n}) \[ F_2(a,b,c) + (-1)^{l_1} F_2(-a,b,c) + (-1)^{l_2} F_2(a,-b,c) + (-1)^{l_3} F_2(a,b,-c) \] \Bigg\},
\qe
where,
\eq
F_1(a,b,c) \= \frac{(a+b+c)^{m_1+m_2+m_3-n+2}(H_{m_1+m_2+m_3-n+2} - \log|a+b+c|)}{(2a)^{m_1+1}(2b)^{m_2+1}(2c)^{m_3+1}} \\
F_2(a,b,c) \= \frac{(a+b+c)^{m_1+m_2+m_3-n+2}\mbox{sign(a+b+c)}}{(2a)^{m_1+1}(2b)^{m_2+1}(2c)^{m_3+1}}.
\qe

These analytic solutions have unfortunately not proved particularly useful in the evaluation of the bispectrum. The solutions that involve large series like above are unstable for large $l$'s and also in the corners of the $\a\b$-triangles, where one of the $k_i$'s tends to zero.  The difficulty with these series for a straightforward numerical implementation is that they generically involve the exact cancellation of many very large terms.

\end{document}